\documentclass[aps,prd,amsmath,floats,floatfix,superscriptaddress,nofootinbib,
twocolumn,
notitlepage,10pt,showpacs,longbibliography,]{revtex4-1}
  
\usepackage{graphicx,color}
\usepackage{amsmath,amssymb}
\usepackage{verbatim}
\usepackage{float}
\usepackage{wasysym}
\usepackage{dsfont}
\usepackage{amsfonts}
\usepackage{mathrsfs}
\usepackage{multirow}
\usepackage{float}
\usepackage{bm}
\usepackage{hyperref}
\usepackage{scalefnt}
\usepackage{xspace}
\usepackage{mathtools}
\usepackage{lipsum} 
\usepackage{physics}
\usepackage{soul}
\hypersetup{
  colorlinks=true,        
  linkcolor=blue,         
  citecolor=cyan,         
}
\usepackage{blindtext, rotating}
\usepackage{pifont}
\usepackage[normalem]{ulem}   
\usepackage{grffile}
\usepackage{booktabs}

\graphicspath{{./}}  

\usepackage{subfigure} %

\usepackage{booktabs, multirow} 
\usepackage{soul}
\usepackage{changepage,threeparttable} 
\usepackage{soul}
\begin{document}


\definecolor{orange}{rgb}{0.9,0.45,0} 
\definecolor{applegreen}{rgb}{0.055, 0.591, 0.0530}
\newcommand{\vjp}[1]{{\textcolor{orange}{[VJ: #1]}}}
\newcommand{\milton}[1]{{\textcolor{cyan}{[Milton: #1]}}}
\newcommand{\dario}[1]{{\textcolor{red}{[Dario: #1]}}}
\newcommand{\juan}[1]{{\textcolor{applegreen}{[Juan: #1]}}}
\newcommand{\argelia}[1]{{\textcolor{magenta}{#1}}}


\title{
Probing a  minimal dark gauge sector via microlensing of compact dark objects}
\author{Juan Barranco}
\affiliation{Departamento de Física, División de Ciencias e Ingenierías, Campus León, Universidad de Guanajuato, C.P. 37150, León, México}
\author{Argelia Bernal}
\affiliation{Departamento de Física, División de Ciencias e Ingenierías, Campus León, Universidad de Guanajuato, C.P. 37150, León, México}
\affiliation{Instituto de F\'isica y Matem\'aticas,
Universidad Michoacana de San Nicol\'as de Hidalgo,
Edificio C-3, Ciudad Universitaria, 58040 Morelia, Michoac\'an, M\'exico}
\author{V\'ictor Jaramillo}
\affiliation{Department of Modern Physics, University of Science and Technology of China, Hefei, Anhui 230026, China}
\author{Dar\'io N\'u\~nez}
\affiliation{Instituto de Ciencias Nucleares, Universidad Nacional
  Aut\'onoma de M\'exico, Circuito Exterior C.U., A.P. 70-543,
  Coyoac\'an, M\'exico 04510, CdMx, M\'exico}
\author{Milton Ruiz}
\affiliation{Departament d'Astronomia i Astrof\'isica, Universitat de València, C/ Dr Moliner 50, 46100, Burjassot (València), Spain}
%

\begin{abstract}
We introduce a minimal Dark Standard Model (DSM) consisting of a single spin-0 particle with dark $U(1)$ gauge symmetry, and completely decoupled  from the visible sector. Characterized only by the scalar mass~$\mu$~and the dark charge~$q$, this framework naturally gives rise to a rich phenomenology,  including stable solitonic configurations that behave as dark ``mini-MACHOs". We numerically build and evolve these gauged  scalar-field solitons, derive their mass-radius relations, and identify a critical charge beyond  which no gravitationally  bound configurations exist. By combining these results with microlensing surveys that exclude compact objects heavier than the asteroid-mass scale ($M\lesssim 10^{-11}M_\odot$), we obtain the constraint $\mu\gtrsim 10\,\rm eV$ for viable configurations, depending on $q$.  Our results represent a step forward in showing that purely gravitational observations can constrain the internal parameters of a dark gauge sector, and provide a framework for exploring broader DSM scenarios through  future probes such as gravitational wave detections.
\end{abstract}
\date{\today}

\maketitle
%
%
\section{Introduction}
\label{Sec:intro}
The discovery of dark matter (DM) through its gravitational interaction with visible matter has been confirmed at galactic \cite{deBlok:2001hbg,Kleyna:2005vw,Walker:2005nt,Battaglia:2008jz} and cosmological scales \cite{WMAP:2003elm,WMAP:2012nax,Planck:2018vyg,ACTPol:2014pbf,SPT:2016izt,DESI:2024kob,DESI:2025zgx}. The 
compelling evidence for the existence of dark matter has driven the development of terrestrial experiments 
pursuing its direct or indirect detection \cite{Gaitskell:2004gd,Gaskins:2016cha}.
Terrestrial experimental searches need to assume a coupling of the DM candidate to visible matter beyond
a pure gravitational interaction in order to get a detection signal. While this experimental approach may 
eventually  detect dark matter, current observations imply that there is no empirical evidence that DM couples 
with standard model (SM) particles~\cite{PandaX-4T:2021bab,LZ:2022lsv,LZ:2024zvo,XENON:2023cxc,PICO:2025rku,TESSERACT:2025tfw,Barman:2022jdg,
Chan:2019ptd,Song:2023xdk,HAWC:2023bti,Li:2025vek}. In this  scenario, an even more elusive possibility arises:
that multiple dark matter particles could exist, none of them coupled to those of the SM. 

We explore the possibility that dark matter may be composed of distinct particle fields preserving local gauge 
symmetries, mediated by their own gauge mediators, in analogy with the Standard Model (SM).
These fields are minimally coupled to gravity but fully decoupled from SM. We refer to this scenario as the Dark Standard Model, DSM. If 
such DSM exists, several questions arise: i) what is its particle content?;
ii) What gauge symmetries dictate the dynamics of DM interactions?; and, most challenging, iii) how could we 
experimentally probe the DSM and its associated dark symmetries?
As a step forward in addressing these questions, we consider a minimal realization of the DSM consisting of a spin-0 
particle with an internal $U(1)$ gauge symmetry.  This represents a textbook example of a gauge theory that has the
lowest number of free parameters: the  scalar mass $\mu$ and the charge $q$. 

Despite having only two parameters, this DSM has a rich phenomenology. For instance, when $q=0$ and for masses 
$\mu$ of the order of $10^{-22}$ eV, the model reduces  to the Scalar Field Dark Matter (SFDM) scenario, also 
known as Fuzzy Dark Matter  Model, which successfully reproduces the Cold Dark Matter (CDM) at cosmological scales 
and, at galactic scales, and may alleviate small-scale (central halo regions) CDM tensions~\cite{Matos:2023usa,Hui:2016ltb}. 

While the SFDM scenario focuses on wave-like dark matter behavior, scalar fields can also behave as particle-like, 
compact configurations. In this scenario, they can also be used within the CDM model, which considers that dark matter is 
composed of non-interactive particles~\cite{Navarro:2008kc}, and these particles can be regarded as scalar-field objects.
Moreover,  microlensing studies~\cite{2012RAA....12..947M,Petac:2022rio,MACHO:1996dlp} 
imply that, if the galactic halos are composed of clusters of massive objects, they cannot have a mass larger than $10^{-11}M_{\odot}$, 
often referred to as mini-MACHOs. Previous studies with scalar field compact objects
(see e.g.~\cite{Hernandez:2004bm,Barranco:2010ib,Barranco:2012ur}) showed that galactic halos could be modeled as 
collisionless ensembles of scalar field MACHOs with masses below than~$ 10^{-7}M_{\odot}$, which was the upper limit at 
that time.

Within this point of view, we show that for
$\mu\simeq 1$~eV, self-gravitating scalar field configurations have masses around 
$10^{-11}M_{\odot}$ and can play the role of  massive compact halo objects (MACHOs). In such a scenario,  the halo 
dynamics would be indistinguishable from that in a CDM halo, since N-body simulations of CDM  are insensitive to particle 
mass granularity smaller than $10^5 M_\odot$ \cite{Navarro:2008kc}. These scalar field mini-MACHOS 
could have formed in the early Universe via a mechanism analogous to the Affleck-Dine Mechanism known in particle 
physics~\cite{Kasuya:2000wx}. Dark neutron stars have recently been investigated within a related framework \cite{Litterer:2025quq}, while Ref.~\cite{Boos:2025nzc} analyzes gravitational microlensing by a class of static, spherically symmetric nonsingular black holes. Related Higgs-portal constructions with a spontaneously broken global U(1) symmetry have also been explored \cite{Ng:2014iqa}. More recently, it has been shown that nonminimal couplings to gravity may produce demagnification signatures in microlensing light curves \cite{Zhang:2025kze}.

Static, spherically symmetric, self-gravitating systems minimally coupled to a scalar field with $U(1)$ 
gauge symmetry, known as charged boson stars,  have been described in~\cite{Jetzer:1989av,Jetzer:1989us,
Pugliese:2013gsa,Lopez:2023phk}.  Recently,  the dynamics of these self-gravitating configurations was studied in~\cite{Jaramillo:2024shi}. In this work, we reinterpret these configurations as 
dark matter structures within the DSM framework. The resulting solutions are everywhere regular and horizonless, 
representing solitonic equilibria of the model. In contrast, black hole configurations with nontrivial scalar 
field profiles exist only for self-interacting scalars under specific resonance conditions~\cite{Herdeiro:2020xmb,
Hong:2020miv}. Such solutions can be continuously connected to the self-interacting (Q-ball–like) boson star sequence 
when a small event horizon is introduced. However,  full 3+1 numerical evolutions show that they develop a non-axisymmetric 
instability~\cite{Nicoules:2025bhh}, while their horizonless counterparts remain stable~\cite{Jaramillo:2024shi}. 
Therefore, within this theory, the viable static gravitational solitons that could serve as MACHO dark matter 
candidates are the Reissner–Nordstr\"om black holes and the gauged boson stars.
Throughout the paper we adopt geometric units ($c=1=G$)  except where stated otherwise. In these units,  the coupling 
parameter $q$ then has units of inverse length and can be normalized by the scalar field mass $\mu$, so that 
ratio $q/\mu$ becomes dimensionless.

%
\section{Gauge dark matter solitons}
The action for the minimal DSM model is
\begin{equation}\label{eq:action}
\begin{split}
S=\int &d^4 x\,\sqrt{-g}\,\left[\frac{1}{16\pi }R-g^{\alpha\nu}\mathcal{D}_\alpha\Phi\mathcal{D}_\nu\Phi^*-\mu^2|\Phi|^2\right.\\
&\left.-\frac{1}{4}F_{\alpha\nu}F^{\alpha\nu}\right] \, ,
\end{split}
\end{equation}
where~$\mu$~is the mass of the spin-0 particle, $F_{\alpha\nu}=\partial_\alpha A_\nu-\partial_\nu A_\alpha$, $g_{\alpha\nu}$ is the 
spacetime metric, $R$ the Ricci scalar. Here  $A_\nu$ is  the  four-potential. Notice that we  use the gauge invariant covariant 
derivative operator defined as $\mathcal{D}_\alpha=\nabla_\alpha+i\,q\,A_\alpha$, where $q$ is the gauge coupling constant or the 
scalar field charge~\cite{Jaramillo:2024shi}.

Variation with respect to $g_{\alpha\nu}$ leads to the Einstein field equations, with the stress energy tensor given by
\begin{equation}
\begin{split}
T_{\alpha\nu}=&2\mathcal{D}_{(\alpha}\Phi \, \mathcal{D}_{\nu)}\Phi^*-g_{\alpha\nu}\left(\mathcal{D}^{\beta}\Phi 
\, \mathcal{D}_{\beta}\Phi^*+\mu^2|\Phi|^2\right) \\
&+F_{\alpha\sigma} F_{\nu\lambda}g^{\sigma\lambda}-\frac{1}{4} g_{\alpha\nu} F_{\sigma\beta} F^{\sigma\beta}.
\label{Eq:tmunu}
\end{split}
\end{equation}
Variation with respect to $A_\alpha$ leads to the Maxwell equations sourced by the spin-0 current $J^\alpha$,
\begin{subequations}
\label{eq:maxwell}
  \begin{eqnarray}
    &&\nabla_\nu F^{\alpha\nu}=J^\alpha:=q J_\Phi^\alpha\,,
    \label{eq:Maxwell0}\\
    &&J^\alpha_\Phi:=i\left({\Phi}^*\mathcal{D}^\alpha\Phi-\Phi \mathcal{D}^\alpha{\Phi}^*\right),\label{eq:current}
  \end{eqnarray}
\end{subequations}
here $J_\Phi^\alpha$ is the conserved current that arises from the U(1) gauge invariance of Eq.~\eqref{eq:action}. 
Finally, variation with respect $\Phi$ leads to the Klein-Gordon equation
\begin{equation}\label{eq:kg}
\mathcal{D}^\alpha \mathcal{D}_\alpha \Phi=\mu^2\Phi \,.
\end{equation}
The minimal energy, self-gravitating realization of  model~in Eq.~\eqref{eq:action} is a static 
spherically symmetric spacetime with a stationary scalar field spherical solutions,\footnote{Often 
refered as charged boson stars~(see~e.g.\cite{Jetzer:1989av}).} characterized by a harmonic time dependence 
of the scalar field of the form $\Phi=\phi(r)e^{i\omega t}$ and an ``electric'' potential $A_\alpha dx^\alpha = V(r)dt$.
We choose coordinates in which the spacetime metric takes the form
\begin{equation}
\label{eq:ansatz_metric}
ds^2 = - e^{2 F_0(r)} dt^2 + e^{2 F_1(r)}\left(dr^2 + r^2 d\Omega^2\right) \,,
\end{equation}
with $d\Omega^2=d\theta^2 + r^2\sin^2\theta\,d\varphi^2$ the solid angle. Notice that the metric is isotropic. 

Under these assumptions for the spacetime and the spin-0 and spin-1 fields, the system reduces to four coupled 
ordinary differential equations for the functions $V(r)$, $\phi(r)$, $F_0(r)$ and $F_1(r)$, see e.g. Appendix 
B~in~\cite{Jaramillo:2024shi} (or \cite{Zhang:2024bjo} for a more general treatment in a non-relativistic limit). We solve the system numerically, noting first that the mass of the scalar field 
$\mu$ can be factored out as the characteristic length scale of the system, and all dimensional quantities rescale 
with appropriate powers~$\mu$. We solve the system using a spectral method with a compactified radial coordinate 
$r$ and a Newton-Raphson algorithm, as described in \cite{Alcubierre:2021psa}, and employ 24 spectral coefficients
for all solutions presented. 
%
%
\begin{figure}
\centering    
\includegraphics[width=0.475\textwidth]{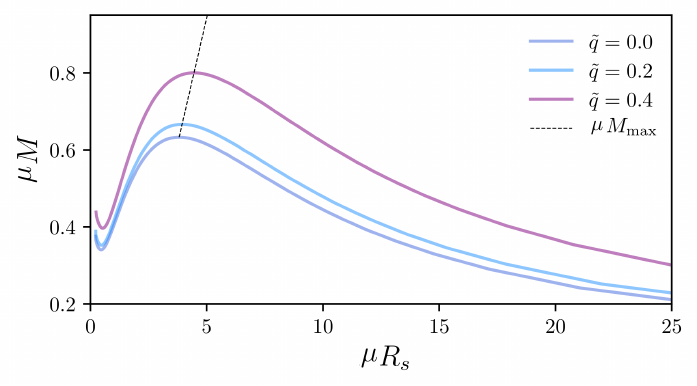}\\
\includegraphics[width=0.475\textwidth]{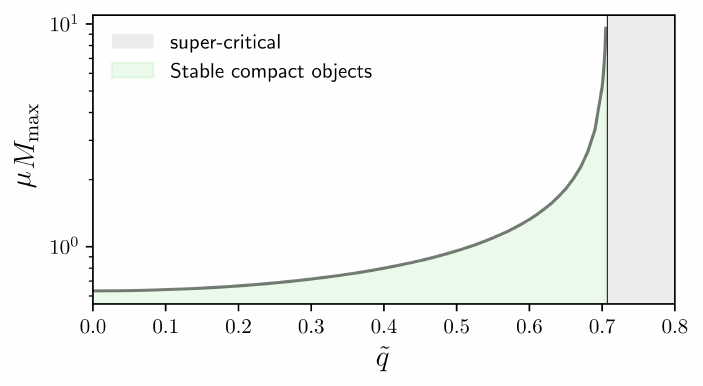}
\caption{Gauge dark matter solitons sequence of solutions. Top panel: Mass vs. radius diagram for families of 
configurations with different values of $q$. Dashed-line displays the maximum mass configurations. Lower panel:
Maximum mass configuration as a function of $q$ in the  region where gravitationally bound, and stable 
configurations exist. Beyond the critical $\tilde{q}_{\rm crit}=1/\sqrt{2}\approx0.707$, no gravitationally 
bound solutions are found.}
\label{fig:BSsequences}
\end{figure}
In the process, the eigenfrequency of the the spin-0 field $\omega$ is determined as part of the solution. 
The equilibrium configurations obtained in this scenario form sequences characterized by the coupling constant $q$,
which has an upper bound of approximately $q=\sqrt{4\pi}\mu$ that is  limited  by Coulomb repulsion \cite{Jetzer:1989av,Pugliese:2013gsa,Lopez:2023phk,Jaramillo:2023lgk,Kleihaus:2009kr,Kumar:2014kna}. Following 
previous literature\cite{Jetzer:1989av,Lopez:2023phk}, we present results in terms of the dimensionless rescaled charge 
$\tilde{q}=q/(\sqrt{8\pi}\mu)$, for which the critical value is $\tilde{q}_{\rm crit}=1/\sqrt{2}$. For the solitons considered here,
the central amplitude of the scalar field parametrizes each sequence of solutions. The top panel of Fig.~\ref{fig:BSsequences} 
shows some sequences for several values of $q$. Here $M$ denotes to the total mass of the configuration, obtained from 
the asymptotic behavior of the metric coefficients:
%
%
\begin{table}
  \centering
    \caption{Gauge dark matter solitons: configurations of maximum mass as a function of the rescaled coupling $\tilde{q}$. 
   We list the dimensionless rescaled charge $\tilde{q}$, the eigenfrequency frequency of the the spin-0 field $\omega$, the 
   maximum value of the scalar field at the origin $\phi(r=0)$, the mass  $\mu M$, the total electrical 
   charge $\mu Q_E$, and the radius $\mu R_s$  of the configurations.
   The critical value $\tilde{q}_{\rm crit}=1/\sqrt{2}\approx0.707$ marks the limit beyond which no gravitationally bound 
  solutions exist. }
  \begin{tabular}{l|cc|c|ccc}
    \hline\hline
    $\tilde{q}$    & $\omega/\mu$ & $\phi(r=0)$    & $\mu M$ & $\mu Q_E$ & $\mu R_s$  \\ \hline
    $0$            & $0.8530$     & $0.0765$       & 0.6330  &   0       & 3.816   \\
    $0.7$          & $0.9958$     & $0.0158$       & 5.203   & 30.96     & 22.86   \\
    $0.705$\quad\, & $0.9987$     & $0.00871$      & 9.567   & 57.69     & 41.71  \\
    \hline\hline
  \end{tabular}
  \label{tab:BSsequences}
\end{table}

\begin{equation}
    M = -\lim_{r\to\infty}\left(r^2\frac{dF_1}{dr}\right) \, ,
\end{equation}
from the $g_{rr}$ component, or 
\begin{equation}
M = \lim_{r\to\infty}\left(r^2\frac{dF_0}{dr}\right)\,,
\end{equation}
from the $g_{tt}$ component. Both should coincide and we use this two quantities to asses accuracy of the code. 
The first definition corresponding to the Arnowitt–Deser–Misner (ADM) mass, while the second corresponds to
the Komar mass (see e.g.~\cite{1979JMP....20..793A,Gourgoulhon:2010ju,Jaramillo:2023lgk}). 
Following~\cite{Alcubierre:2021psa},
the radius of the  configuration is defined as
%
the areal radius $R$ where $C:=M(R)/R=e^{-F_1}M(r)/r$ is maximized. Here $M(r)$ denotes the Misner-Sharp function, which in isotropic coordinates is $M(r)=-r^2e^{F_1}\partial_rF_1\left(1+r\partial_r F_1 / 2\right)$.
The conserved charge $Q$ associated with the conserved current $J_\Phi^\alpha$ in 
Eq.~\eqref{eq:current}, is given by 
%
%
\begin{figure}
\centering    
\includegraphics[width=0.49\textwidth]{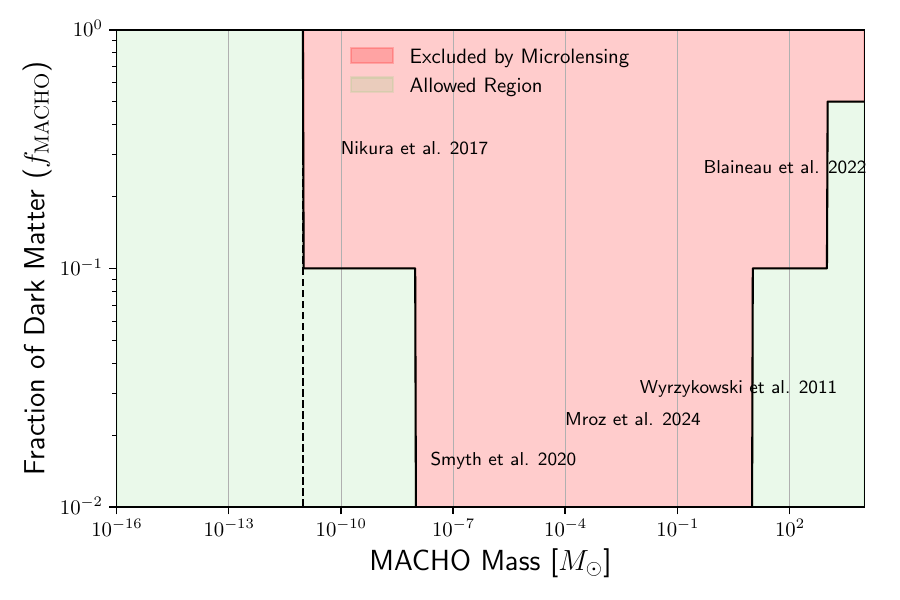}
\caption{Summary of current microlensing limits on the MACHO dark matter fraction as a function of lens mass. The solid black line marks the combined 95\% confidence level exclusion boundary obtained from recent surveys. Niikura et al.~\cite{Niikura:2017zjd} constrain the sub-lunar window around $10^{-11}-10^{-6}\,M_\odot$, Smyth et al.\cite{Smyth:2019whb} probe the Earth–Jupiter range $10^{-6}-10^{-3}\,M_\odot$, Wyrzykowski et al.~\cite{Wyrzykowski:2011tr} constrain the stellar regime $\sim10^{-1}-10\,M_\odot$, and Blaineau et al.~\cite{Blaineau:2022nhy} cover the intermediate range $10-10^{3}\,M_\odot$. Updated OGLE analyses combining the results of Refs.~\cite{Mroz:2024mse,Mroz:2024wia} further impose global limits, excluding MACHOs contributing more than 1\% of the dark matter in the mass range $10^{-8}-10\,M_\odot$, and more than 10\% in the broader interval $10^{-9}-10^{3}\,M_\odot$. The red-shaded regions above the exclusion boundary are therefore ruled out.}
\label{fig:microlensing}
\end{figure}
\begin{equation}
Q=4\pi\intop_{0}^{\infty}e^{3F_1-F_0}r^{2}\phi^{2}\left(Vq+\omega\right)\,dr\,,
\label{eq:Npart}
\end{equation}
which is related to the total ``electrical" charge $Q_E$ of the soliton through~$Q_E=q\,Q$. Far from the source,
the electric potential decays as 
\begin{equation}
V(r)=-\frac{Q_E}{4\pi r}+\mathcal{O}\left(\frac{1}{r^2}\right)\,, 
\end{equation}
which we again use to check accuracy of our numerical solutions. We have checked that the relative differences between
the ADM and Komar masses, as well as between the two charge definitions, are below $10^{-4}$. 

Note that the total charge $Q$ has no definite sign, as it depends on the product of the gauge coupling $q$, which is
fixed in sign, and $\omega$, which  can be either  positive or negative. 
The Maxwell Eqs. given by~Eq.~\eqref{eq:maxwell} is invariant under $(\omega,A_\alpha)\rightarrow(-\omega,-A_\alpha)$, implying that
once the parameter $q$ is fixed, a change in the sign of $\omega$ reverses the total electrical charge, as seen from
Eq.~\eqref{eq:Npart}.

Fig.~\ref{fig:BSsequences} shows that, for each value of $q$,  there is a maximum value of $M$ (dashed-line
in the top panel)
that increases with $q$. Table~\ref{tab:BSsequences} lists additional properties of the corresponding
dark-matter solitons, including their radius $R_s$, and total charge $Q_E$.   Stability analysis, both 
perturbative~\cite{Jetzer:1989us} and non-perturbative~\cite{Lopez:2023phk,Jaramillo:2024shi},  have concluded 
that, configurations with  radii larger than that of the maximum mass configurations are stable. Therefore, all 
configurations on the right side of the maximum (see top panel of Fig.~\ref{fig:BSsequences}) remain stable, 
while those to the left side are 
unstable: they either collapse to a black hole, relax into a lower-mass equilibrium state, or disperse.
The lower panel of Fig.~\ref{fig:BSsequences} shows $M_{\rm max}$ as a function of the coupling parameter $q$ 
in the range $[0,0.705]$. Configurations beyond the critical charge value $\tilde{q}_{\rm crit}=1/\sqrt{2}\approx 
0.707$ were found to be gravitationally unbound and therefore unstable. In summary, for each value of $q$, there exist
a sequence of solutions. If $q<q_{\rm crit}$, a global maximum-mass configuration separates the stable from the
unstable branch. These numerical results constrain the dimensionless quantity $\mu M$ as function of $q$ 
(see lower panel of Fig.~\ref{fig:BSsequences}).

In the next section, we  use microlensing limits on MACHOs to further constrain the allowed values of $M$, and 
consequently the parameters~$\mu$ and $q$.

%
\section{Constraints on the free~$\mu$ and~$q$ from microlensing limits}
%
%
\begin{figure}
\centering    
\includegraphics[width=0.475\textwidth]{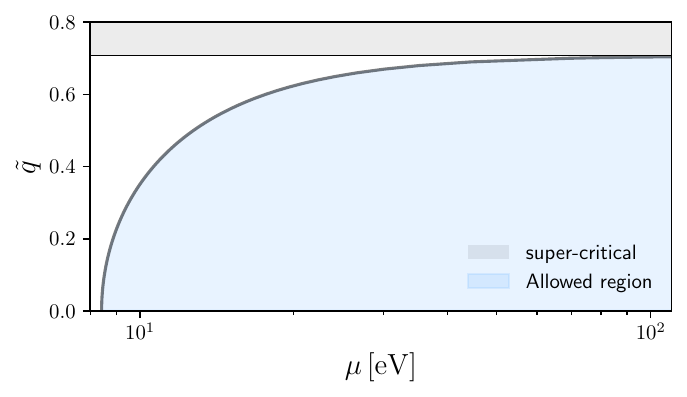}
\includegraphics[width=0.475\textwidth]{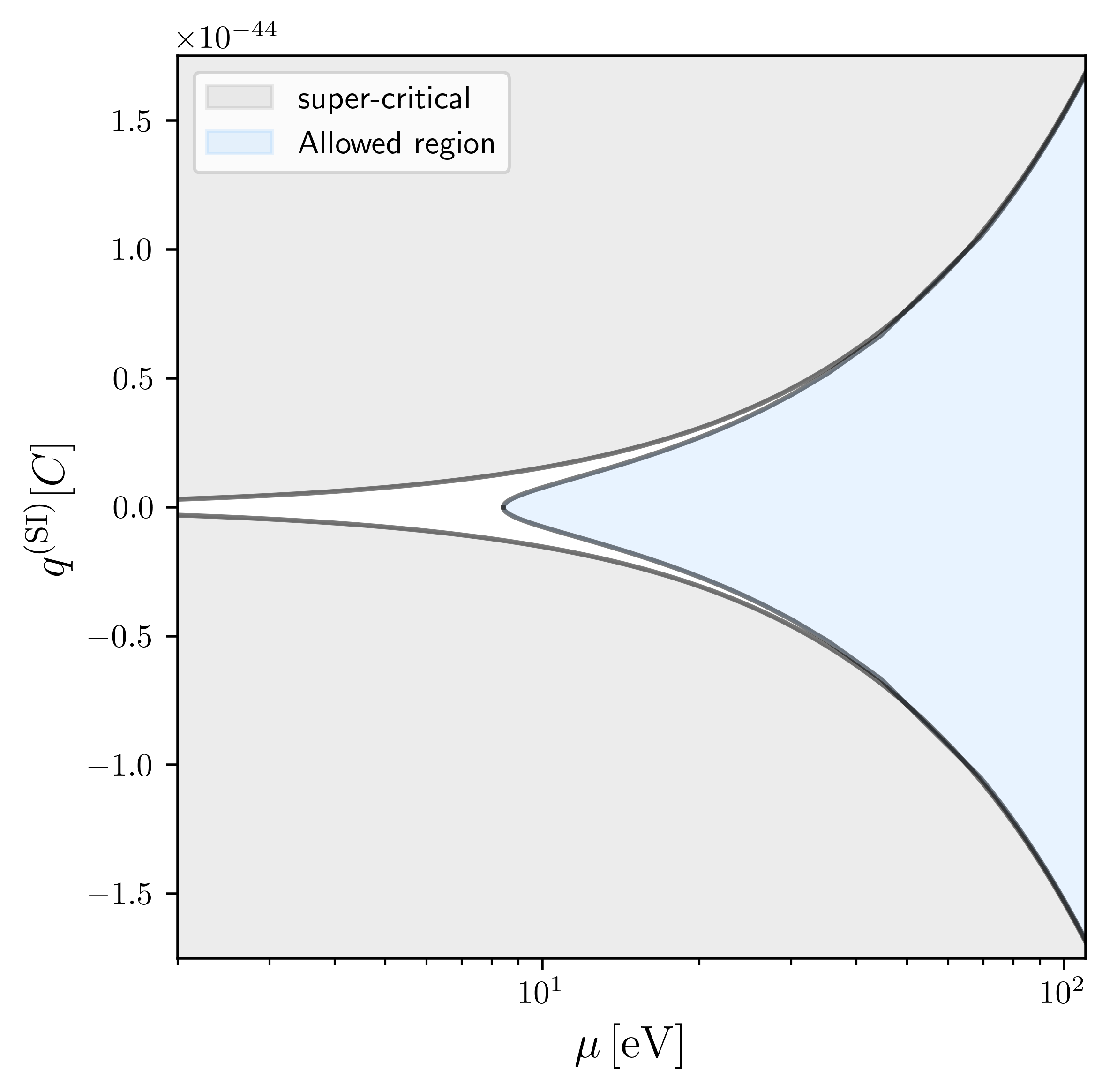}
\caption{Constraints in the plane $(\mu,q)$ parameter space derived from the non-observation of microlensing 
events. The shaded region indicates the values of the scalar-field mass $\mu$ and coupling $q$
consistent with current MACHO limits. Larger masses are excluded by Eq.~\eqref{eq:constraint}.}
\label{fig:mu}
\end{figure}
Current microlensing observations have imposed tight constraints on the fraction of dark matter that may be due 
to MACHOs. As summarized in Fig.~\ref{fig:microlensing}, these limits depend on the mass of the compact objects 
$\mathcal{M}$. Microlensing observations have effectively excluded MACHOs with $\mathcal{M} > 10^{-11}M_\odot$ 
as the dominant dark matter. In the sub-lunar to Earth-mass range ($10^{-11} \lesssim \mathcal{M}/M_\odot 
\lesssim 10^{-6}$), the Subaru Hyper Suprime-Cam (HSC) survey of M31 by Niikura et al.~\cite{Niikura:2017zjd} 
detected only one candidate event, placing strong upper limits on the  MACHO fraction of dark matter 
and effectively closing the so-called ``lunar-mass window''.
At lower masses~($10^{-16} \lesssim \mathcal{M}/M_\odot \lesssim 10^{-10}$), Smyth et al.~\cite{Smyth:2019whb} 
reanalyzed the same HSC data, incorporating finite-source-size effects. They found that these effects significantly 
weaken constraints at the smallest masses, suggesting that asteroid-mass  MACHOs could still constitute all of the 
dark matter. At solar masses and above, OGLE observations toward the Magellanic Clouds, particularly those by 
Wyrzykowski et al.~\cite{Wyrzykowski:2011tr}, show that the detected microlensing events are consistent with known 
stellar populations, thereby ruling out MACHOs with $\mathcal{M} \sim 0.1-10\,M_\odot$ as more than a small fraction 
of the dark halo. For heavier objects in the intermediate range ($\mathcal{M} \sim 10 - 10^3\,M_\odot$), Blaineau et al.
\cite{Blaineau:2022nhy} combined MACHO and EROS datasets, finding that such black hole–like MACHOs can account for 
at most about $15\%-50\%$ of the dark matter, depending on mass. Overall, microlensing surveys have largely ruled 
out MACHOs as the dominant form of dark matter, leaving room primarily for mini-MACHOs  in the asteroid-mass regime.

We model MACHOs using the gauge dark matter solitons described in the previous section and show how 
microlensing limits constrain the parameters $\mu$ and $q$. For a fixed $q$, we numerically obtain  a maximum value 
of $\mu M$, a consequence of the self-gravitating nature of the solitons in the minimal gauge dark matter mode in
Eq.~\eqref{eq:action}~\cite{Kaup:1968zz}.

From the above observational constraints, MACHO masses must satisfy
\begin{equation}\label{eq:asteroids}
    \mathcal{M}\lesssim10^{-11}M_\odot = 1.48\times10^{-8} {\rm m} \, .
\end{equation}
Therefore, we require that the maximum mass of the solitonic configurations in Eq.\eqref{eq:action} not exceed 
this bound, i.e.
 \begin{equation}
 M_{\rm max} < 1.48 \times 10^{-8}\, {\rm m}\,.
 \label{eq:constraint}
 \end{equation}
 Otherwise, stable solutions exceeding this bound would have masses already  ruled out by current microlensing 
 limits. This condition imposes a constraint on the scalar-field mass~$\mu$:
\begin{equation}
    \mu > 6.76\times10^{7}x(q)\, {\rm m}^{-1}=13.3 x(q){\rm eV}
\end{equation}
where $x(q)=\mu M_{\rm max}$  is obtained from the numerical sequences shown in the lower panel of 
Fig.~\ref{fig:BSsequences}, or from some specific configurations in Table~\ref{tab:BSsequences}.
The corresponding constraints on $\mu$ as function of $q$ are displayed in the top panel of Fig.~\ref{fig:mu}. 
In the lower panel, we also show the resulting allowed region with the coupling constant given 
in SI units, using the conversion
\begin{equation}
    q^{({\rm SI})} = \frac{\sqrt{8\pi G}\hbar}{c^2\sqrt{\mu_0}}\, \mu\, \tilde{q} = 
    2.167\times10^{-46} \, \left(\frac{\mu}{\mathrm{eV}}\right) \tilde{q}\, \mathrm{C} \,,
\end{equation}
where $G$ is Newton's gravitational constant, $\hbar$ the reduced Planck constant, 
$c$ the speed of light in vacuum, and $\mu_0$ the vacuum permeability. 

The above results identify a narrow region in the parameter space where gravitationally bound, stable 
solitons remain compatible with current microlensing constraints. In particular, microlensing non-detections 
restrict the dark scalar sector to relatively heavy field masses and small gauge couplings, defining a narrow 
window of viable mini-MACHO configurations within the DSM framework.

%
\section{Conclusions}
The dynamics of visible matter on both galactic and cosmological scales 
have implications for the existence
of dark matter through their gravitational interaction. To date, there is no experimental evidence
that dark matter interacts with visible matter through any force other than gravity. However, it
is possible that the dark matter sector is not composed of a single particle, but instead forms
a self-consistent dark analogue of the Standard Model with its own gauge interactions.

We introduced a minimal Dark Standard Model with a dark $U(1)$ gauge symmetry, giving rise to
solitonic ``mini-MACHOs." We have presented this scalar field model not as fuzzy dark matter, but
within the CDM framework, where dark matter consists of non-interactive particles. We showed that
these particles could be composed of a scalar field, whose rest mass must be greater than 
$10\,\rm eV$, with more restrictive constraints for a non-zero dark charge.

By combining our derived mass-radius relations for these solitons with microlensing constraints on
asteroid-mass objects, we obtain a lower bound of $\gtrsim 10\,\rm eV$ on the scalar mass $\mu$.
We derived this bound from the non-observation of microlensing events, combined with the general
properties of gauged scalar field stars. Our main results are summarized in Fig.~\ref{fig:mu},
where, for each value of the gauge coupling $q$, we identify a corresponding minimum allowed scalar-field
mass. In particular, for $q = 0$, the bound is of order $10\,\rm eV$.

From Fig.~\ref{fig:mu}, one may conclude that current microlensing observations impose a lower limit
on the scalar-field mass $\mu$, while the gauge coupling $q$ is restricted by the theoretical condition
$\tilde{q}<\tilde{q}_{\rm crit}=1/\sqrt{2}$, which ensures the existence of gravitationally bound
equilibrium configurations. Only sufficiently heavy fields can form compact objects compatible with
current surveys. This shifts the viable parameter space of the minimal DSM to $\mu$-values well above
the ultralight fuzzy/axion-like regime.

These configurations are everywhere regular and horizonless, forming dynamically stable solitonic
equilibria of the model in Eq.~\eqref{eq:action}. In contrast, as mentioned above, black hole counterparts
with nontrivial scalar hair arise only in self-interacting theories under specific resonance conditions
and are unstable. Therefore, within the considered framework, the viable static gravitational solitons
consistent with current astrophysical constraints are the Reissner-Nordstr\"om black holes and the gauged
boson stars, the latter standing as natural, stable MACHO dark matter candidates. The limits we have
derived on the scalar field mass could also be affected by considering additional terms in the scalar
potential, such as a self-interaction.

The dark matter model proposed in this work also opens new directions for exploring its potential observational
signatures. Collisions or mergers of these dark asteroids could contribute to a stochastic gravitational
wave background with distinctive spectral features, as discussed previously~(see e.g.~\cite{Clesse:2016vqa,Carr:2016drx,
Giudice:2016zpa,Jaramillo:2022zwg, Lira:2024cma}). Furthermore, the accretion of such dark compact objects
onto black holes could endow Reissner–Nordstr\"om and Kerr-Newman geometries with an effective dark charge,
providing a novel physical interpretation of the charge parameter in these solutions. This perspective could
be applied in other contexts where the conventional view has been used, such as to matter in the vicinity
of black holes. This possibility motivates further studies of geodesic motion and observable effects in
such spacetimes, following the approaches developed in~\cite{Herrera-Aguilar:2015kea, Wang:2022ouq, Villegas:2022jzt}.

Within this scenario, there are multiple other possibilities that can be studied with different gauge mediators
and dark matter particles. These bounds are expected to hold in more general dark-gauge sectors with additional
fields or interactions. Given the opening of new observational tools and experiments that probe the Universe
through gravitational interactions, such as gravitational wave observations, it now seems possible to begin
the systematic study of a dark matter standard model along the lines described in this work, potentially
connecting gravitational phenomena to the internal structure of the dark sector.
%

\acknowledgments
This work has been partially supported by the  National Key R\&D Program of China under grant 
No.~2022YFC2204603, by the Generalitat Valenciana (grants CIDEGENT/2021/046  and Prometeo 
CIPROM/2022/49), and by the Spanish Agencia Estatal de Investigación  (grants PID2024-159689NB-C21 
funded by MCIN/AEI/10.13039/501100011033, PRE2019-087617,  and ERDF A way of making Europe). We acknowledge computational resources and technical sup-925
port of the Spanish Supercomputing Network through the926
use of MareNostrum at the Barcelona Supercomputing Cen-927
ter (AECT-2023-1-0006).
\bibliography{ref} 

\begin{thebibliography}{68}%
\makeatletter
\providecommand \@ifxundefined [1]{%
 \@ifx{#1\undefined}
}%
\providecommand \@ifnum [1]{%
 \ifnum #1\expandafter \@firstoftwo
 \else \expandafter \@secondoftwo
 \fi
}%
\providecommand \@ifx [1]{%
 \ifx #1\expandafter \@firstoftwo
 \else \expandafter \@secondoftwo
 \fi
}%
\providecommand \natexlab [1]{#1}%
\providecommand \enquote  [1]{``#1''}%
\providecommand \bibnamefont  [1]{#1}%
\providecommand \bibfnamefont [1]{#1}%
\providecommand \citenamefont [1]{#1}%
\providecommand \href@noop [0]{\@secondoftwo}%
\providecommand \href [0]{\begingroup \@sanitize@url \@href}%
\providecommand \@href[1]{\@@startlink{#1}\@@href}%
\providecommand \@@href[1]{\endgroup#1\@@endlink}%
\providecommand \@sanitize@url [0]{\catcode `\\12\catcode `\$12\catcode `\&12\catcode `\#12\catcode `\^12\catcode `\_12\catcode `\%12\relax}%
\providecommand \@@startlink[1]{}%
\providecommand \@@endlink[0]{}%
\providecommand \url  [0]{\begingroup\@sanitize@url \@url }%
\providecommand \@url [1]{\endgroup\@href {#1}{\urlprefix }}%
\providecommand \urlprefix  [0]{URL }%
\providecommand \Eprint [0]{\href }%
\providecommand \doibase [0]{http://dx.doi.org/}%
\providecommand \selectlanguage [0]{\@gobble}%
\providecommand \bibinfo  [0]{\@secondoftwo}%
\providecommand \bibfield  [0]{\@secondoftwo}%
\providecommand \translation [1]{[#1]}%
\providecommand \BibitemOpen [0]{}%
\providecommand \bibitemStop [0]{}%
\providecommand \bibitemNoStop [0]{.\EOS\space}%
\providecommand \EOS [0]{\spacefactor3000\relax}%
\providecommand \BibitemShut  [1]{\csname bibitem#1\endcsname}%
\let\auto@bib@innerbib\@empty
\bibitem [{\citenamefont {de~Blok}\ \emph {et~al.}(2001)\citenamefont {de~Blok}, \citenamefont {McGaugh}, \citenamefont {Bosma},\ and\ \citenamefont {Rubin}}]{deBlok:2001hbg}%
  \BibitemOpen
  \bibfield  {author} {\bibinfo {author} {\bibfnamefont {W.~J.~G.}\ \bibnamefont {de~Blok}}, \bibinfo {author} {\bibfnamefont {Stacy~S.}\ \bibnamefont {McGaugh}}, \bibinfo {author} {\bibfnamefont {Albert}\ \bibnamefont {Bosma}}, \ and\ \bibinfo {author} {\bibfnamefont {Vera~C.}\ \bibnamefont {Rubin}},\ }\bibfield  {title} {\enquote {\bibinfo {title} {{Mass density profiles of LSB galaxies}},}\ }\href {\doibase 10.1086/320262} {\bibfield  {journal} {\bibinfo  {journal} {Astrophys. J. Lett.}\ }\textbf {\bibinfo {volume} {552}},\ \bibinfo {pages} {L23--L26} (\bibinfo {year} {2001})},\ \Eprint {http://arxiv.org/abs/astro-ph/0103102} {arXiv:astro-ph/0103102} \BibitemShut {NoStop}%
\bibitem [{\citenamefont {Kleyna}\ \emph {et~al.}(2005)\citenamefont {Kleyna}, \citenamefont {Wilkinson}, \citenamefont {Evans},\ and\ \citenamefont {Gilmore}}]{Kleyna:2005vw}%
  \BibitemOpen
  \bibfield  {author} {\bibinfo {author} {\bibfnamefont {Jan~T.}\ \bibnamefont {Kleyna}}, \bibinfo {author} {\bibfnamefont {Mark~I.}\ \bibnamefont {Wilkinson}}, \bibinfo {author} {\bibfnamefont {N.~Wyn}\ \bibnamefont {Evans}}, \ and\ \bibinfo {author} {\bibfnamefont {Gerard}\ \bibnamefont {Gilmore}},\ }\bibfield  {title} {\enquote {\bibinfo {title} {{Ursa Major: A Missing low-mass CDM halo?}}}\ }\href {\doibase 10.1086/491654} {\bibfield  {journal} {\bibinfo  {journal} {Astrophys. J. Lett.}\ }\textbf {\bibinfo {volume} {630}},\ \bibinfo {pages} {L141--L144} (\bibinfo {year} {2005})},\ \Eprint {http://arxiv.org/abs/astro-ph/0507154} {arXiv:astro-ph/0507154} \BibitemShut {NoStop}%
\bibitem [{\citenamefont {Walker}\ \emph {et~al.}(2006)\citenamefont {Walker}, \citenamefont {Mateo}, \citenamefont {Olszewski}, \citenamefont {Bernstein}, \citenamefont {Wang},\ and\ \citenamefont {Woodroofe}}]{Walker:2005nt}%
  \BibitemOpen
  \bibfield  {author} {\bibinfo {author} {\bibfnamefont {Matthew~G.}\ \bibnamefont {Walker}}, \bibinfo {author} {\bibfnamefont {Mario}\ \bibnamefont {Mateo}}, \bibinfo {author} {\bibfnamefont {Edward~W.}\ \bibnamefont {Olszewski}}, \bibinfo {author} {\bibfnamefont {Rebecca~A.}\ \bibnamefont {Bernstein}}, \bibinfo {author} {\bibfnamefont {Xiao}\ \bibnamefont {Wang}}, \ and\ \bibinfo {author} {\bibfnamefont {Michael}\ \bibnamefont {Woodroofe}},\ }\bibfield  {title} {\enquote {\bibinfo {title} {{Internal kinematics of the fornax dwarf spheroidal galaxy}},}\ }\href {\doibase 10.1086/500193} {\bibfield  {journal} {\bibinfo  {journal} {Astron. J.}\ }\textbf {\bibinfo {volume} {131}},\ \bibinfo {pages} {2114--2139} (\bibinfo {year} {2006})},\ \bibinfo {note} {[Erratum: Astron.J. 132, 968--968 (2006)]},\ \Eprint {http://arxiv.org/abs/astro-ph/0511465} {arXiv:astro-ph/0511465} \BibitemShut {NoStop}%
\bibitem [{\citenamefont {Battaglia}\ \emph {et~al.}(2008)\citenamefont {Battaglia}, \citenamefont {Helmi}, \citenamefont {Tolstoy}, \citenamefont {Irwin}, \citenamefont {Hill},\ and\ \citenamefont {Jablonka}}]{Battaglia:2008jz}%
  \BibitemOpen
  \bibfield  {author} {\bibinfo {author} {\bibfnamefont {G.}~\bibnamefont {Battaglia}}, \bibinfo {author} {\bibfnamefont {A.}~\bibnamefont {Helmi}}, \bibinfo {author} {\bibfnamefont {E.}~\bibnamefont {Tolstoy}}, \bibinfo {author} {\bibfnamefont {M.}~\bibnamefont {Irwin}}, \bibinfo {author} {\bibfnamefont {V.}~\bibnamefont {Hill}}, \ and\ \bibinfo {author} {\bibfnamefont {P.}~\bibnamefont {Jablonka}},\ }\bibfield  {title} {\enquote {\bibinfo {title} {{The kinematic status and mass content of the Sculptor dwarf spheroidal galaxy}},}\ }\href {\doibase 10.1086/590179} {\bibfield  {journal} {\bibinfo  {journal} {Astrophys. J. Lett.}\ }\textbf {\bibinfo {volume} {681}},\ \bibinfo {pages} {L13} (\bibinfo {year} {2008})},\ \Eprint {http://arxiv.org/abs/0802.4220} {arXiv:0802.4220 [astro-ph]} \BibitemShut {NoStop}%
\bibitem [{\citenamefont {Spergel}\ \emph {et~al.}(2003)\citenamefont {Spergel} \emph {et~al.}}]{WMAP:2003elm}%
  \BibitemOpen
  \bibfield  {author} {\bibinfo {author} {\bibfnamefont {D.~N.}\ \bibnamefont {Spergel}} \emph {et~al.} (\bibinfo {collaboration} {WMAP}),\ }\bibfield  {title} {\enquote {\bibinfo {title} {{First year Wilkinson Microwave Anisotropy Probe (WMAP) observations: Determination of cosmological parameters}},}\ }\href {\doibase 10.1086/377226} {\bibfield  {journal} {\bibinfo  {journal} {Astrophys. J. Suppl.}\ }\textbf {\bibinfo {volume} {148}},\ \bibinfo {pages} {175--194} (\bibinfo {year} {2003})},\ \Eprint {http://arxiv.org/abs/astro-ph/0302209} {arXiv:astro-ph/0302209} \BibitemShut {NoStop}%
\bibitem [{\citenamefont {Hinshaw}\ \emph {et~al.}(2013)\citenamefont {Hinshaw} \emph {et~al.}}]{WMAP:2012nax}%
  \BibitemOpen
  \bibfield  {author} {\bibinfo {author} {\bibfnamefont {G.}~\bibnamefont {Hinshaw}} \emph {et~al.} (\bibinfo {collaboration} {WMAP}),\ }\bibfield  {title} {\enquote {\bibinfo {title} {{Nine-Year Wilkinson Microwave Anisotropy Probe (WMAP) Observations: Cosmological Parameter Results}},}\ }\href {\doibase 10.1088/0067-0049/208/2/19} {\bibfield  {journal} {\bibinfo  {journal} {Astrophys. J. Suppl.}\ }\textbf {\bibinfo {volume} {208}},\ \bibinfo {pages} {19} (\bibinfo {year} {2013})},\ \Eprint {http://arxiv.org/abs/1212.5226} {arXiv:1212.5226 [astro-ph.CO]} \BibitemShut {NoStop}%
\bibitem [{\citenamefont {Aghanim}\ \emph {et~al.}(2020)\citenamefont {Aghanim} \emph {et~al.}}]{Planck:2018vyg}%
  \BibitemOpen
  \bibfield  {author} {\bibinfo {author} {\bibfnamefont {N.}~\bibnamefont {Aghanim}} \emph {et~al.} (\bibinfo {collaboration} {Planck}),\ }\bibfield  {title} {\enquote {\bibinfo {title} {{Planck 2018 results. VI. Cosmological parameters}},}\ }\href {\doibase 10.1051/0004-6361/201833910} {\bibfield  {journal} {\bibinfo  {journal} {Astron. Astrophys.}\ }\textbf {\bibinfo {volume} {641}},\ \bibinfo {pages} {A6} (\bibinfo {year} {2020})},\ \bibinfo {note} {[Erratum: Astron.Astrophys. 652, C4 (2021)]},\ \Eprint {http://arxiv.org/abs/1807.06209} {arXiv:1807.06209 [astro-ph.CO]} \BibitemShut {NoStop}%
\bibitem [{\citenamefont {Naess}\ \emph {et~al.}(2014)\citenamefont {Naess} \emph {et~al.}}]{ACTPol:2014pbf}%
  \BibitemOpen
  \bibfield  {author} {\bibinfo {author} {\bibfnamefont {Sigurd}\ \bibnamefont {Naess}} \emph {et~al.} (\bibinfo {collaboration} {ACTPol}),\ }\bibfield  {title} {\enquote {\bibinfo {title} {{The Atacama Cosmology Telescope: CMB Polarization at $200<\ell<9000$}},}\ }\href {\doibase 10.1088/1475-7516/2014/10/007} {\bibfield  {journal} {\bibinfo  {journal} {JCAP}\ }\textbf {\bibinfo {volume} {10}},\ \bibinfo {pages} {007} (\bibinfo {year} {2014})},\ \Eprint {http://arxiv.org/abs/1405.5524} {arXiv:1405.5524 [astro-ph.CO]} \BibitemShut {NoStop}%
\bibitem [{\citenamefont {de~Haan}\ \emph {et~al.}(2016)\citenamefont {de~Haan} \emph {et~al.}}]{SPT:2016izt}%
  \BibitemOpen
  \bibfield  {author} {\bibinfo {author} {\bibfnamefont {T.}~\bibnamefont {de~Haan}} \emph {et~al.} (\bibinfo {collaboration} {SPT}),\ }\bibfield  {title} {\enquote {\bibinfo {title} {{Cosmological Constraints from Galaxy Clusters in the 2500 square-degree SPT-SZ Survey}},}\ }\href {\doibase 10.3847/0004-637X/832/1/95} {\bibfield  {journal} {\bibinfo  {journal} {Astrophys. J.}\ }\textbf {\bibinfo {volume} {832}},\ \bibinfo {pages} {95} (\bibinfo {year} {2016})},\ \Eprint {http://arxiv.org/abs/1603.06522} {arXiv:1603.06522 [astro-ph.CO]} \BibitemShut {NoStop}%
\bibitem [{\citenamefont {Lodha}\ \emph {et~al.}(2025)\citenamefont {Lodha} \emph {et~al.}}]{DESI:2024kob}%
  \BibitemOpen
  \bibfield  {author} {\bibinfo {author} {\bibfnamefont {K.}~\bibnamefont {Lodha}} \emph {et~al.} (\bibinfo {collaboration} {DESI}),\ }\bibfield  {title} {\enquote {\bibinfo {title} {{DESI 2024: Constraints on physics-focused aspects of dark energy using DESI DR1 BAO data}},}\ }\href {\doibase 10.1103/PhysRevD.111.023532} {\bibfield  {journal} {\bibinfo  {journal} {Phys. Rev. D}\ }\textbf {\bibinfo {volume} {111}},\ \bibinfo {pages} {023532} (\bibinfo {year} {2025})},\ \Eprint {http://arxiv.org/abs/2405.13588} {arXiv:2405.13588 [astro-ph.CO]} \BibitemShut {NoStop}%
\bibitem [{\citenamefont {Abdul~Karim}\ \emph {et~al.}(2025)\citenamefont {Abdul~Karim} \emph {et~al.}}]{DESI:2025zgx}%
  \BibitemOpen
  \bibfield  {author} {\bibinfo {author} {\bibfnamefont {M.}~\bibnamefont {Abdul~Karim}} \emph {et~al.} (\bibinfo {collaboration} {DESI}),\ }\bibfield  {title} {\enquote {\bibinfo {title} {{DESI DR2 Results II: Measurements of Baryon Acoustic Oscillations and Cosmological Constraints}},}\ }\href@noop {} {\  (\bibinfo {year} {2025})},\ \Eprint {http://arxiv.org/abs/2503.14738} {arXiv:2503.14738 [astro-ph.CO]} \BibitemShut {NoStop}%
\bibitem [{\citenamefont {Gaitskell}(2004)}]{Gaitskell:2004gd}%
  \BibitemOpen
  \bibfield  {author} {\bibinfo {author} {\bibfnamefont {R.~J.}\ \bibnamefont {Gaitskell}},\ }\bibfield  {title} {\enquote {\bibinfo {title} {{Direct detection of dark matter}},}\ }\href {\doibase 10.1146/annurev.nucl.54.070103.181244} {\bibfield  {journal} {\bibinfo  {journal} {Ann. Rev. Nucl. Part. Sci.}\ }\textbf {\bibinfo {volume} {54}},\ \bibinfo {pages} {315--359} (\bibinfo {year} {2004})}\BibitemShut {NoStop}%
\bibitem [{\citenamefont {Gaskins}(2016)}]{Gaskins:2016cha}%
  \BibitemOpen
  \bibfield  {author} {\bibinfo {author} {\bibfnamefont {Jennifer~M.}\ \bibnamefont {Gaskins}},\ }\bibfield  {title} {\enquote {\bibinfo {title} {{A review of indirect searches for particle dark matter}},}\ }\href {\doibase 10.1080/00107514.2016.1175160} {\bibfield  {journal} {\bibinfo  {journal} {Contemp. Phys.}\ }\textbf {\bibinfo {volume} {57}},\ \bibinfo {pages} {496--525} (\bibinfo {year} {2016})},\ \Eprint {http://arxiv.org/abs/1604.00014} {arXiv:1604.00014 [astro-ph.HE]} \BibitemShut {NoStop}%
\bibitem [{\citenamefont {Meng}\ \emph {et~al.}(2021)\citenamefont {Meng} \emph {et~al.}}]{PandaX-4T:2021bab}%
  \BibitemOpen
  \bibfield  {author} {\bibinfo {author} {\bibfnamefont {Yue}\ \bibnamefont {Meng}} \emph {et~al.} (\bibinfo {collaboration} {PandaX-4T}),\ }\bibfield  {title} {\enquote {\bibinfo {title} {{Dark Matter Search Results from the PandaX-4T Commissioning Run}},}\ }\href {\doibase 10.1103/PhysRevLett.127.261802} {\bibfield  {journal} {\bibinfo  {journal} {Phys. Rev. Lett.}\ }\textbf {\bibinfo {volume} {127}},\ \bibinfo {pages} {261802} (\bibinfo {year} {2021})},\ \Eprint {http://arxiv.org/abs/2107.13438} {arXiv:2107.13438 [hep-ex]} \BibitemShut {NoStop}%
\bibitem [{\citenamefont {Aalbers}\ \emph {et~al.}(2023)\citenamefont {Aalbers} \emph {et~al.}}]{LZ:2022lsv}%
  \BibitemOpen
  \bibfield  {author} {\bibinfo {author} {\bibfnamefont {J.}~\bibnamefont {Aalbers}} \emph {et~al.} (\bibinfo {collaboration} {LZ}),\ }\bibfield  {title} {\enquote {\bibinfo {title} {{First Dark Matter Search Results from the LUX-ZEPLIN (LZ) Experiment}},}\ }\href {\doibase 10.1103/PhysRevLett.131.041002} {\bibfield  {journal} {\bibinfo  {journal} {Phys. Rev. Lett.}\ }\textbf {\bibinfo {volume} {131}},\ \bibinfo {pages} {041002} (\bibinfo {year} {2023})},\ \Eprint {http://arxiv.org/abs/2207.03764} {arXiv:2207.03764 [hep-ex]} \BibitemShut {NoStop}%
\bibitem [{\citenamefont {Aalbers}\ \emph {et~al.}(2024)\citenamefont {Aalbers} \emph {et~al.}}]{LZ:2024zvo}%
  \BibitemOpen
  \bibfield  {author} {\bibinfo {author} {\bibfnamefont {J.}~\bibnamefont {Aalbers}} \emph {et~al.} (\bibinfo {collaboration} {LZ}),\ }\bibfield  {title} {\enquote {\bibinfo {title} {{Dark Matter Search Results from 4.2 Tonne-Years of Exposure of the LUX-ZEPLIN (LZ) Experiment}},}\ }\href@noop {} {\  (\bibinfo {year} {2024})},\ \Eprint {http://arxiv.org/abs/2410.17036} {arXiv:2410.17036 [hep-ex]} \BibitemShut {NoStop}%
\bibitem [{\citenamefont {Aprile}\ \emph {et~al.}(2023)\citenamefont {Aprile} \emph {et~al.}}]{XENON:2023cxc}%
  \BibitemOpen
  \bibfield  {author} {\bibinfo {author} {\bibfnamefont {E.}~\bibnamefont {Aprile}} \emph {et~al.} (\bibinfo {collaboration} {XENON}),\ }\bibfield  {title} {\enquote {\bibinfo {title} {{First Dark Matter Search with Nuclear Recoils from the XENONnT Experiment}},}\ }\href {\doibase 10.1103/PhysRevLett.131.041003} {\bibfield  {journal} {\bibinfo  {journal} {Phys. Rev. Lett.}\ }\textbf {\bibinfo {volume} {131}},\ \bibinfo {pages} {041003} (\bibinfo {year} {2023})},\ \Eprint {http://arxiv.org/abs/2303.14729} {arXiv:2303.14729 [hep-ex]} \BibitemShut {NoStop}%
\bibitem [{\citenamefont {Adams}\ \emph {et~al.}(2025)\citenamefont {Adams} \emph {et~al.}}]{PICO:2025rku}%
  \BibitemOpen
  \bibfield  {author} {\bibinfo {author} {\bibfnamefont {E.}~\bibnamefont {Adams}} \emph {et~al.} (\bibinfo {collaboration} {PICO}),\ }\bibfield  {title} {\enquote {\bibinfo {title} {{Absorption of Fermionic Dark Matter in the PICO-60 C$_{3}$F$_{8}$ Bubble Chamber}},}\ }\href@noop {} {\  (\bibinfo {year} {2025})},\ \Eprint {http://arxiv.org/abs/2504.13089} {arXiv:2504.13089 [hep-ex]} \BibitemShut {NoStop}%
\bibitem [{\citenamefont {Chang}\ \emph {et~al.}(2025)\citenamefont {Chang} \emph {et~al.}}]{TESSERACT:2025tfw}%
  \BibitemOpen
  \bibfield  {author} {\bibinfo {author} {\bibfnamefont {C.~L.}\ \bibnamefont {Chang}} \emph {et~al.} (\bibinfo {collaboration} {TESSERACT}),\ }\bibfield  {title} {\enquote {\bibinfo {title} {{First Limits on Light Dark Matter Interactions in a Low Threshold Two Channel Athermal Phonon Detector from the TESSERACT Collaboration}},}\ }\href@noop {} {\  (\bibinfo {year} {2025})},\ \Eprint {http://arxiv.org/abs/2503.03683} {arXiv:2503.03683 [hep-ex]} \BibitemShut {NoStop}%
\bibitem [{\citenamefont {Barman}\ \emph {et~al.}(2023)\citenamefont {Barman}, \citenamefont {B\'elanger}, \citenamefont {Bhattacherjee}, \citenamefont {Godbole},\ and\ \citenamefont {Sengupta}}]{Barman:2022jdg}%
  \BibitemOpen
  \bibfield  {author} {\bibinfo {author} {\bibfnamefont {Rahool~Kumar}\ \bibnamefont {Barman}}, \bibinfo {author} {\bibfnamefont {Genevi\`eve}\ \bibnamefont {B\'elanger}}, \bibinfo {author} {\bibfnamefont {Biplob}\ \bibnamefont {Bhattacherjee}}, \bibinfo {author} {\bibfnamefont {Rohini~M.}\ \bibnamefont {Godbole}}, \ and\ \bibinfo {author} {\bibfnamefont {Rhitaja}\ \bibnamefont {Sengupta}},\ }\bibfield  {title} {\enquote {\bibinfo {title} {{Is Light Neutralino Thermal Dark Matter in the Phenomenological Minimal Supersymmetric Standard Model Ruled Out?}}}\ }\href {\doibase 10.1103/PhysRevLett.131.011802} {\bibfield  {journal} {\bibinfo  {journal} {Phys. Rev. Lett.}\ }\textbf {\bibinfo {volume} {131}},\ \bibinfo {pages} {011802} (\bibinfo {year} {2023})},\ \Eprint {http://arxiv.org/abs/2207.06238} {arXiv:2207.06238 [hep-ph]} \BibitemShut {NoStop}%
\bibitem [{\citenamefont {Chan}\ \emph {et~al.}(2019)\citenamefont {Chan}, \citenamefont {Cui}, \citenamefont {Liu},\ and\ \citenamefont {Leung}}]{Chan:2019ptd}%
  \BibitemOpen
  \bibfield  {author} {\bibinfo {author} {\bibfnamefont {Man~Ho}\ \bibnamefont {Chan}}, \bibinfo {author} {\bibfnamefont {Lang}\ \bibnamefont {Cui}}, \bibinfo {author} {\bibfnamefont {Jun}\ \bibnamefont {Liu}}, \ and\ \bibinfo {author} {\bibfnamefont {Chun~Sing}\ \bibnamefont {Leung}},\ }\bibfield  {title} {\enquote {\bibinfo {title} {{Ruling out $\sim 100-300$ GeV thermal relic annihilating dark matter by radio observation of the Andromeda galaxy}},}\ }\href {\doibase 10.3847/1538-4357/aafe0b} {\bibfield  {journal} {\bibinfo  {journal} {Astrophys. J.}\ }\textbf {\bibinfo {volume} {872}},\ \bibinfo {pages} {177} (\bibinfo {year} {2019})},\ \Eprint {http://arxiv.org/abs/1901.04638} {arXiv:1901.04638 [astro-ph.GA]} \BibitemShut {NoStop}%
\bibitem [{\citenamefont {Song}\ \emph {et~al.}(2024)\citenamefont {Song}, \citenamefont {Murase},\ and\ \citenamefont {Kheirandish}}]{Song:2023xdk}%
  \BibitemOpen
  \bibfield  {author} {\bibinfo {author} {\bibfnamefont {Deheng}\ \bibnamefont {Song}}, \bibinfo {author} {\bibfnamefont {Kohta}\ \bibnamefont {Murase}}, \ and\ \bibinfo {author} {\bibfnamefont {Ali}\ \bibnamefont {Kheirandish}},\ }\bibfield  {title} {\enquote {\bibinfo {title} {{Constraining decaying very heavy dark matter from galaxy clusters with 14 year Fermi-LAT data}},}\ }\href {\doibase 10.1088/1475-7516/2024/03/024} {\bibfield  {journal} {\bibinfo  {journal} {JCAP}\ }\textbf {\bibinfo {volume} {03}},\ \bibinfo {pages} {024} (\bibinfo {year} {2024})},\ \Eprint {http://arxiv.org/abs/2308.00589} {arXiv:2308.00589 [astro-ph.HE]} \BibitemShut {NoStop}%
\bibitem [{\citenamefont {Albert}\ \emph {et~al.}(2024)\citenamefont {Albert} \emph {et~al.}}]{HAWC:2023bti}%
  \BibitemOpen
  \bibfield  {author} {\bibinfo {author} {\bibfnamefont {A.}~\bibnamefont {Albert}} \emph {et~al.} (\bibinfo {collaboration} {HAWC}),\ }\bibfield  {title} {\enquote {\bibinfo {title} {{Search for decaying dark matter in the Virgo cluster of galaxies with HAWC}},}\ }\href {\doibase 10.1103/PhysRevD.109.043034} {\bibfield  {journal} {\bibinfo  {journal} {Phys. Rev. D}\ }\textbf {\bibinfo {volume} {109}},\ \bibinfo {pages} {043034} (\bibinfo {year} {2024})},\ \Eprint {http://arxiv.org/abs/2309.03973} {arXiv:2309.03973 [astro-ph.HE]} \BibitemShut {NoStop}%
\bibitem [{\citenamefont {Li}\ and\ \citenamefont {Han}(2025)}]{Li:2025vek}%
  \BibitemOpen
  \bibfield  {author} {\bibinfo {author} {\bibfnamefont {Shang}\ \bibnamefont {Li}}\ and\ \bibinfo {author} {\bibfnamefont {Feng}\ \bibnamefont {Han}},\ }\bibfield  {title} {\enquote {\bibinfo {title} {{Search for dark matter annihilation to \ensuremath{\gamma}-rays from nearby galaxy clusters with Fermi-LAT data}},}\ }\href {\doibase 10.1093/mnras/staf636} {\bibfield  {journal} {\bibinfo  {journal} {Mon. Not. Roy. Astron. Soc.}\ }\textbf {\bibinfo {volume} {539}},\ \bibinfo {pages} {2242--2247} (\bibinfo {year} {2025})}\BibitemShut {NoStop}%
\bibitem [{\citenamefont {Matos}\ \emph {et~al.}(2024)\citenamefont {Matos}, \citenamefont {Ure\~na L\'opez},\ and\ \citenamefont {Lee}}]{Matos:2023usa}%
  \BibitemOpen
  \bibfield  {author} {\bibinfo {author} {\bibfnamefont {Tonatiuh}\ \bibnamefont {Matos}}, \bibinfo {author} {\bibfnamefont {Luis~A.}\ \bibnamefont {Ure\~na L\'opez}}, \ and\ \bibinfo {author} {\bibfnamefont {Jae-Weon}\ \bibnamefont {Lee}},\ }\bibfield  {title} {\enquote {\bibinfo {title} {{Short review of the main achievements of the scalar field, fuzzy, ultralight, wave, BEC dark matter model}},}\ }\href {\doibase 10.3389/fspas.2024.1347518} {\bibfield  {journal} {\bibinfo  {journal} {Front. Astron. Space Sci.}\ }\textbf {\bibinfo {volume} {11}},\ \bibinfo {pages} {1347518} (\bibinfo {year} {2024})},\ \Eprint {http://arxiv.org/abs/2312.00254} {arXiv:2312.00254 [astro-ph.CO]} \BibitemShut {NoStop}%
\bibitem [{\citenamefont {Hui}\ \emph {et~al.}(2017)\citenamefont {Hui}, \citenamefont {Ostriker}, \citenamefont {Tremaine},\ and\ \citenamefont {Witten}}]{Hui:2016ltb}%
  \BibitemOpen
  \bibfield  {author} {\bibinfo {author} {\bibfnamefont {Lam}\ \bibnamefont {Hui}}, \bibinfo {author} {\bibfnamefont {Jeremiah~P.}\ \bibnamefont {Ostriker}}, \bibinfo {author} {\bibfnamefont {Scott}\ \bibnamefont {Tremaine}}, \ and\ \bibinfo {author} {\bibfnamefont {Edward}\ \bibnamefont {Witten}},\ }\bibfield  {title} {\enquote {\bibinfo {title} {{Ultralight scalars as cosmological dark matter}},}\ }\href {\doibase 10.1103/PhysRevD.95.043541} {\bibfield  {journal} {\bibinfo  {journal} {Phys. Rev. D}\ }\textbf {\bibinfo {volume} {95}},\ \bibinfo {pages} {043541} (\bibinfo {year} {2017})},\ \Eprint {http://arxiv.org/abs/1610.08297} {arXiv:1610.08297 [astro-ph.CO]} \BibitemShut {NoStop}%
\bibitem [{\citenamefont {Navarro}\ \emph {et~al.}(2010)\citenamefont {Navarro}, \citenamefont {Ludlow}, \citenamefont {Springel}, \citenamefont {Wang}, \citenamefont {Vogelsberger}, \citenamefont {White}, \citenamefont {Jenkins}, \citenamefont {Frenk},\ and\ \citenamefont {Helmi}}]{Navarro:2008kc}%
  \BibitemOpen
  \bibfield  {author} {\bibinfo {author} {\bibfnamefont {Julio~F.}\ \bibnamefont {Navarro}}, \bibinfo {author} {\bibfnamefont {Aaron}\ \bibnamefont {Ludlow}}, \bibinfo {author} {\bibfnamefont {Volker}\ \bibnamefont {Springel}}, \bibinfo {author} {\bibfnamefont {Jie}\ \bibnamefont {Wang}}, \bibinfo {author} {\bibfnamefont {Mark}\ \bibnamefont {Vogelsberger}}, \bibinfo {author} {\bibfnamefont {Simon D.~M.}\ \bibnamefont {White}}, \bibinfo {author} {\bibfnamefont {Adrian}\ \bibnamefont {Jenkins}}, \bibinfo {author} {\bibfnamefont {Carlos~S.}\ \bibnamefont {Frenk}}, \ and\ \bibinfo {author} {\bibfnamefont {Amina}\ \bibnamefont {Helmi}},\ }\bibfield  {title} {\enquote {\bibinfo {title} {{The Diversity and Similarity of Cold Dark Matter Halos}},}\ }\href {\doibase 10.1111/j.1365-2966.2009.15878.x} {\bibfield  {journal} {\bibinfo  {journal} {Mon. Not. Roy. Astron. Soc.}\ }\textbf {\bibinfo {volume} {402}},\ \bibinfo {pages} {21} (\bibinfo {year} {2010})},\ \Eprint {http://arxiv.org/abs/0810.1522} {arXiv:0810.1522
  [astro-ph]} \BibitemShut {NoStop}%
\bibitem [{\citenamefont {{Mao}}(2012)}]{2012RAA....12..947M}%
  \BibitemOpen
  \bibfield  {author} {\bibinfo {author} {\bibfnamefont {Shude}\ \bibnamefont {{Mao}}},\ }\bibfield  {title} {\enquote {\bibinfo {title} {{Astrophysical applications of gravitational microlensing}},}\ }\href {\doibase 10.1088/1674-4527/12/8/005} {\bibfield  {journal} {\bibinfo  {journal} {Research in Astronomy and Astrophysics}\ }\textbf {\bibinfo {volume} {12}},\ \bibinfo {pages} {947--972} (\bibinfo {year} {2012})},\ \Eprint {http://arxiv.org/abs/1207.3720} {arXiv:1207.3720 [astro-ph.GA]} \BibitemShut {NoStop}%
\bibitem [{\citenamefont {Peta{\v{c}}}\ \emph {et~al.}(2022)\citenamefont {Peta{\v{c}}}, \citenamefont {Lavalle},\ and\ \citenamefont {Jedamzik}}]{Petac:2022rio}%
  \BibitemOpen
  \bibfield  {author} {\bibinfo {author} {\bibfnamefont {Mihael}\ \bibnamefont {Peta{\v{c}}}}, \bibinfo {author} {\bibfnamefont {Julien}\ \bibnamefont {Lavalle}}, \ and\ \bibinfo {author} {\bibfnamefont {Karsten}\ \bibnamefont {Jedamzik}},\ }\bibfield  {title} {\enquote {\bibinfo {title} {{Microlensing constraints on clustered primordial black holes}},}\ }\href {\doibase 10.1103/PhysRevD.105.083520} {\bibfield  {journal} {\bibinfo  {journal} {Phys. Rev. D}\ }\textbf {\bibinfo {volume} {105}},\ \bibinfo {pages} {083520} (\bibinfo {year} {2022})},\ \Eprint {http://arxiv.org/abs/2201.02521} {arXiv:2201.02521 [astro-ph.CO]} \BibitemShut {NoStop}%
\bibitem [{\citenamefont {Alcock}\ \emph {et~al.}(1996)\citenamefont {Alcock} \emph {et~al.}}]{MACHO:1996dlp}%
  \BibitemOpen
  \bibfield  {author} {\bibinfo {author} {\bibfnamefont {C.}~\bibnamefont {Alcock}} \emph {et~al.} (\bibinfo {collaboration} {MACHO}),\ }\bibfield  {title} {\enquote {\bibinfo {title} {{The MACHO project: limits on planetary mass dark matter in the galactic halo from gravitational microlensing}},}\ }\href {\doibase 10.1086/178005} {\bibfield  {journal} {\bibinfo  {journal} {Astrophys. J.}\ }\textbf {\bibinfo {volume} {471}},\ \bibinfo {pages} {774} (\bibinfo {year} {1996})},\ \Eprint {http://arxiv.org/abs/astro-ph/9604176} {arXiv:astro-ph/9604176} \BibitemShut {NoStop}%
\bibitem [{\citenamefont {Hernandez}\ \emph {et~al.}(2004)\citenamefont {Hernandez}, \citenamefont {Matos}, \citenamefont {Sussman},\ and\ \citenamefont {Verbin}}]{Hernandez:2004bm}%
  \BibitemOpen
  \bibfield  {author} {\bibinfo {author} {\bibfnamefont {Xavier}\ \bibnamefont {Hernandez}}, \bibinfo {author} {\bibfnamefont {Tonatiuh}\ \bibnamefont {Matos}}, \bibinfo {author} {\bibfnamefont {Roberto~A.}\ \bibnamefont {Sussman}}, \ and\ \bibinfo {author} {\bibfnamefont {Yosef}\ \bibnamefont {Verbin}},\ }\bibfield  {title} {\enquote {\bibinfo {title} {{Scalar field mini-machos: A New explanation for galactic dark matter}},}\ }\href {\doibase 10.1103/PhysRevD.70.043537} {\bibfield  {journal} {\bibinfo  {journal} {Phys. Rev. D}\ }\textbf {\bibinfo {volume} {70}},\ \bibinfo {pages} {043537} (\bibinfo {year} {2004})},\ \Eprint {http://arxiv.org/abs/astro-ph/0407245} {arXiv:astro-ph/0407245} \BibitemShut {NoStop}%
\bibitem [{\citenamefont {Barranco}\ and\ \citenamefont {Bernal}(2011)}]{Barranco:2010ib}%
  \BibitemOpen
  \bibfield  {author} {\bibinfo {author} {\bibfnamefont {J.}~\bibnamefont {Barranco}}\ and\ \bibinfo {author} {\bibfnamefont {A.}~\bibnamefont {Bernal}},\ }\bibfield  {title} {\enquote {\bibinfo {title} {{Self-gravitating system made of axions}},}\ }\href {\doibase 10.1103/PhysRevD.83.043525} {\bibfield  {journal} {\bibinfo  {journal} {Phys. Rev. D}\ }\textbf {\bibinfo {volume} {83}},\ \bibinfo {pages} {043525} (\bibinfo {year} {2011})},\ \Eprint {http://arxiv.org/abs/1001.1769} {arXiv:1001.1769 [astro-ph.CO]} \BibitemShut {NoStop}%
\bibitem [{\citenamefont {Barranco}\ \emph {et~al.}(2013)\citenamefont {Barranco}, \citenamefont {Monteverde},\ and\ \citenamefont {Delepine}}]{Barranco:2012ur}%
  \BibitemOpen
  \bibfield  {author} {\bibinfo {author} {\bibfnamefont {J.}~\bibnamefont {Barranco}}, \bibinfo {author} {\bibfnamefont {A.~Carrillo}\ \bibnamefont {Monteverde}}, \ and\ \bibinfo {author} {\bibfnamefont {D.}~\bibnamefont {Delepine}},\ }\bibfield  {title} {\enquote {\bibinfo {title} {{Can the dark matter halo be a collisionless ensemble of axion stars?}}}\ }\href {\doibase 10.1103/PhysRevD.87.103011} {\bibfield  {journal} {\bibinfo  {journal} {Phys. Rev. D}\ }\textbf {\bibinfo {volume} {87}},\ \bibinfo {pages} {103011} (\bibinfo {year} {2013})},\ \Eprint {http://arxiv.org/abs/1212.2254} {arXiv:1212.2254 [astro-ph.CO]} \BibitemShut {NoStop}%
\bibitem [{\citenamefont {Kasuya}\ and\ \citenamefont {Kawasaki}(2000)}]{Kasuya:2000wx}%
  \BibitemOpen
  \bibfield  {author} {\bibinfo {author} {\bibfnamefont {S.}~\bibnamefont {Kasuya}}\ and\ \bibinfo {author} {\bibfnamefont {M.}~\bibnamefont {Kawasaki}},\ }\bibfield  {title} {\enquote {\bibinfo {title} {{Q Ball formation in the gravity mediated SUSY breaking scenario}},}\ }\href {\doibase 10.1103/PhysRevD.62.023512} {\bibfield  {journal} {\bibinfo  {journal} {Phys. Rev. D}\ }\textbf {\bibinfo {volume} {62}},\ \bibinfo {pages} {023512} (\bibinfo {year} {2000})},\ \Eprint {http://arxiv.org/abs/hep-ph/0002285} {arXiv:hep-ph/0002285} \BibitemShut {NoStop}%
\bibitem [{\citenamefont {Litterer}\ and\ \citenamefont {Rosa}(2025)}]{Litterer:2025quq}%
  \BibitemOpen
  \bibfield  {author} {\bibinfo {author} {\bibfnamefont {Jacob~A.}\ \bibnamefont {Litterer}}\ and\ \bibinfo {author} {\bibfnamefont {Jo{\~a}o~G.}\ \bibnamefont {Rosa}},\ }\bibfield  {title} {\enquote {\bibinfo {title} {{Dark neutron stars from a heavy dark sector}},}\ }\href@noop {} {\  (\bibinfo {year} {2025})},\ \Eprint {http://arxiv.org/abs/2511.01984} {arXiv:2511.01984 [hep-ph]} \BibitemShut {NoStop}%
\bibitem [{\citenamefont {Boos}\ and\ \citenamefont {Hu}(2026)}]{Boos:2025nzc}%
  \BibitemOpen
  \bibfield  {author} {\bibinfo {author} {\bibfnamefont {Jens}\ \bibnamefont {Boos}}\ and\ \bibinfo {author} {\bibfnamefont {Hao}\ \bibnamefont {Hu}},\ }\bibfield  {title} {\enquote {\bibinfo {title} {{Microlensing of nonsingular black holes at finite size: A ray tracing approach}},}\ }\href {\doibase 10.1103/dv28-xm9x} {\bibfield  {journal} {\bibinfo  {journal} {Phys. Rev. D}\ }\textbf {\bibinfo {volume} {113}},\ \bibinfo {pages} {024065} (\bibinfo {year} {2026})},\ \Eprint {http://arxiv.org/abs/2510.10282} {arXiv:2510.10282 [gr-qc]} \BibitemShut {NoStop}%
\bibitem [{\citenamefont {Ng}\ \emph {et~al.}(2014)\citenamefont {Ng}, \citenamefont {Tu},\ and\ \citenamefont {Yuan}}]{Ng:2014iqa}%
  \BibitemOpen
  \bibfield  {author} {\bibinfo {author} {\bibfnamefont {Kin-Wang}\ \bibnamefont {Ng}}, \bibinfo {author} {\bibfnamefont {Huitzu}\ \bibnamefont {Tu}}, \ and\ \bibinfo {author} {\bibfnamefont {Tzu-Chiang}\ \bibnamefont {Yuan}},\ }\bibfield  {title} {\enquote {\bibinfo {title} {{Dark photons as fractional cosmic neutrino masquerader}},}\ }\href {\doibase 10.1088/1475-7516/2014/09/035} {\bibfield  {journal} {\bibinfo  {journal} {JCAP}\ }\textbf {\bibinfo {volume} {09}},\ \bibinfo {pages} {035} (\bibinfo {year} {2014})},\ \Eprint {http://arxiv.org/abs/1406.1993} {arXiv:1406.1993 [hep-ph]} \BibitemShut {NoStop}%
\bibitem [{\citenamefont {Zhang}(2025{\natexlab{a}})}]{Zhang:2025kze}%
  \BibitemOpen
  \bibfield  {author} {\bibinfo {author} {\bibfnamefont {Hong-Yi}\ \bibnamefont {Zhang}},\ }\bibfield  {title} {\enquote {\bibinfo {title} {{Demagnifying gravitational lenses as probes of dark matter structures and nonminimal couplings to gravity}},}\ }\href@noop {} {\  (\bibinfo {year} {2025}{\natexlab{a}})},\ \Eprint {http://arxiv.org/abs/2510.05575} {arXiv:2510.05575 [gr-qc]} \BibitemShut {NoStop}%
\bibitem [{\citenamefont {Jetzer}\ and\ \citenamefont {van~der Bij}(1989)}]{Jetzer:1989av}%
  \BibitemOpen
  \bibfield  {author} {\bibinfo {author} {\bibfnamefont {P.}~\bibnamefont {Jetzer}}\ and\ \bibinfo {author} {\bibfnamefont {J.~J.}\ \bibnamefont {van~der Bij}},\ }\bibfield  {title} {\enquote {\bibinfo {title} {{CHARGED BOSON STARS}},}\ }\href {\doibase 10.1016/0370-2693(89)90941-6} {\bibfield  {journal} {\bibinfo  {journal} {Phys. Lett. B}\ }\textbf {\bibinfo {volume} {227}},\ \bibinfo {pages} {341--346} (\bibinfo {year} {1989})}\BibitemShut {NoStop}%
\bibitem [{\citenamefont {Jetzer}(1989)}]{Jetzer:1989us}%
  \BibitemOpen
  \bibfield  {author} {\bibinfo {author} {\bibfnamefont {P.}~\bibnamefont {Jetzer}},\ }\bibfield  {title} {\enquote {\bibinfo {title} {{Stability of Charged Boson Stars}},}\ }\href {\doibase 10.1016/0370-2693(89)90689-8} {\bibfield  {journal} {\bibinfo  {journal} {Phys. Lett. B}\ }\textbf {\bibinfo {volume} {231}},\ \bibinfo {pages} {433--438} (\bibinfo {year} {1989})}\BibitemShut {NoStop}%
\bibitem [{\citenamefont {Pugliese}\ \emph {et~al.}(2013)\citenamefont {Pugliese}, \citenamefont {Quevedo}, \citenamefont {Rueda~H.},\ and\ \citenamefont {Ruffini}}]{Pugliese:2013gsa}%
  \BibitemOpen
  \bibfield  {author} {\bibinfo {author} {\bibfnamefont {Daniela}\ \bibnamefont {Pugliese}}, \bibinfo {author} {\bibfnamefont {Hernando}\ \bibnamefont {Quevedo}}, \bibinfo {author} {\bibfnamefont {Jorge~A.}\ \bibnamefont {Rueda~H.}}, \ and\ \bibinfo {author} {\bibfnamefont {Remo}\ \bibnamefont {Ruffini}},\ }\bibfield  {title} {\enquote {\bibinfo {title} {{On charged boson stars}},}\ }\href {\doibase 10.1103/PhysRevD.88.024053} {\bibfield  {journal} {\bibinfo  {journal} {Phys. Rev. D}\ }\textbf {\bibinfo {volume} {88}},\ \bibinfo {pages} {024053} (\bibinfo {year} {2013})},\ \Eprint {http://arxiv.org/abs/1305.4241} {arXiv:1305.4241 [astro-ph.HE]} \BibitemShut {NoStop}%
\bibitem [{\citenamefont {L\'opez}\ and\ \citenamefont {Alcubierre}(2023)}]{Lopez:2023phk}%
  \BibitemOpen
  \bibfield  {author} {\bibinfo {author} {\bibfnamefont {Jos\'e~Dami\'an}\ \bibnamefont {L\'opez}}\ and\ \bibinfo {author} {\bibfnamefont {Miguel}\ \bibnamefont {Alcubierre}},\ }\bibfield  {title} {\enquote {\bibinfo {title} {{Charged boson stars revisited}},}\ }\href {\doibase 10.1007/s10714-023-03113-8} {\bibfield  {journal} {\bibinfo  {journal} {Gen. Rel. Grav.}\ }\textbf {\bibinfo {volume} {55}},\ \bibinfo {pages} {72} (\bibinfo {year} {2023})},\ \Eprint {http://arxiv.org/abs/2303.04066} {arXiv:2303.04066 [gr-qc]} \BibitemShut {NoStop}%
\bibitem [{\citenamefont {Jaramillo}\ \emph {et~al.}(2025)\citenamefont {Jaramillo}, \citenamefont {N\'u\~nez}, \citenamefont {Ruiz},\ and\ \citenamefont {Zilh\~ao}}]{Jaramillo:2024shi}%
  \BibitemOpen
  \bibfield  {author} {\bibinfo {author} {\bibfnamefont {V\'\i{}ctor}\ \bibnamefont {Jaramillo}}, \bibinfo {author} {\bibfnamefont {Dar\'\i{}o}\ \bibnamefont {N\'u\~nez}}, \bibinfo {author} {\bibfnamefont {Milton}\ \bibnamefont {Ruiz}}, \ and\ \bibinfo {author} {\bibfnamefont {Miguel}\ \bibnamefont {Zilh\~ao}},\ }\bibfield  {title} {\enquote {\bibinfo {title} {{Full 3D nonlinear dynamics of charged and magnetized boson stars}},}\ }\href {\doibase 10.1103/PhysRevD.111.024070} {\bibfield  {journal} {\bibinfo  {journal} {Phys. Rev. D}\ }\textbf {\bibinfo {volume} {111}},\ \bibinfo {pages} {024070} (\bibinfo {year} {2025})},\ \Eprint {http://arxiv.org/abs/2411.07284} {arXiv:2411.07284 [gr-qc]} \BibitemShut {NoStop}%
\bibitem [{\citenamefont {Herdeiro}\ and\ \citenamefont {Radu}(2020)}]{Herdeiro:2020xmb}%
  \BibitemOpen
  \bibfield  {author} {\bibinfo {author} {\bibfnamefont {Carlos A.~R.}\ \bibnamefont {Herdeiro}}\ and\ \bibinfo {author} {\bibfnamefont {Eugen}\ \bibnamefont {Radu}},\ }\bibfield  {title} {\enquote {\bibinfo {title} {{Spherical electro-vacuum black holes with resonant, scalar $Q$-hair}},}\ }\href {\doibase 10.1140/epjc/s10052-020-7976-9} {\bibfield  {journal} {\bibinfo  {journal} {Eur. Phys. J. C}\ }\textbf {\bibinfo {volume} {80}},\ \bibinfo {pages} {390} (\bibinfo {year} {2020})},\ \Eprint {http://arxiv.org/abs/2004.00336} {arXiv:2004.00336 [gr-qc]} \BibitemShut {NoStop}%
\bibitem [{\citenamefont {Hong}\ \emph {et~al.}(2020)\citenamefont {Hong}, \citenamefont {Suzuki},\ and\ \citenamefont {Yamada}}]{Hong:2020miv}%
  \BibitemOpen
  \bibfield  {author} {\bibinfo {author} {\bibfnamefont {Jeong-Pyong}\ \bibnamefont {Hong}}, \bibinfo {author} {\bibfnamefont {Motoo}\ \bibnamefont {Suzuki}}, \ and\ \bibinfo {author} {\bibfnamefont {Masaki}\ \bibnamefont {Yamada}},\ }\bibfield  {title} {\enquote {\bibinfo {title} {{Spherically Symmetric Scalar Hair for Charged Black Holes}},}\ }\href {\doibase 10.1103/PhysRevLett.125.111104} {\bibfield  {journal} {\bibinfo  {journal} {Phys. Rev. Lett.}\ }\textbf {\bibinfo {volume} {125}},\ \bibinfo {pages} {111104} (\bibinfo {year} {2020})},\ \Eprint {http://arxiv.org/abs/2004.03148} {arXiv:2004.03148 [gr-qc]} \BibitemShut {NoStop}%
\bibitem [{\citenamefont {Nicoules}\ \emph {et~al.}(2025)\citenamefont {Nicoules}, \citenamefont {Ferreira}, \citenamefont {Herdeiro}, \citenamefont {Radu},\ and\ \citenamefont {Zilh{\~a}o}}]{Nicoules:2025bhh}%
  \BibitemOpen
  \bibfield  {author} {\bibinfo {author} {\bibfnamefont {Jordan}\ \bibnamefont {Nicoules}}, \bibinfo {author} {\bibfnamefont {Jos{\'e}}\ \bibnamefont {Ferreira}}, \bibinfo {author} {\bibfnamefont {Carlos A.~R.}\ \bibnamefont {Herdeiro}}, \bibinfo {author} {\bibfnamefont {Eugen}\ \bibnamefont {Radu}}, \ and\ \bibinfo {author} {\bibfnamefont {Miguel}\ \bibnamefont {Zilh{\~a}o}},\ }\bibfield  {title} {\enquote {\bibinfo {title} {{Splitting the Gravitational Atom: Instabilities of Black Holes with Synchronized/Resonant Hair}},}\ }\href@noop {} {\  (\bibinfo {year} {2025})},\ \Eprint {http://arxiv.org/abs/2509.20450} {arXiv:2509.20450 [gr-qc]} \BibitemShut {NoStop}%
\bibitem [{\citenamefont {Zhang}(2025{\natexlab{b}})}]{Zhang:2024bjo}%
  \BibitemOpen
  \bibfield  {author} {\bibinfo {author} {\bibfnamefont {Hong-Yi}\ \bibnamefont {Zhang}},\ }\bibfield  {title} {\enquote {\bibinfo {title} {{Unified view of scalar and vector dark matter solitons}},}\ }\href {\doibase 10.1007/JHEP04(2025)174} {\bibfield  {journal} {\bibinfo  {journal} {JHEP}\ }\textbf {\bibinfo {volume} {04}},\ \bibinfo {pages} {174} (\bibinfo {year} {2025}{\natexlab{b}})},\ \Eprint {http://arxiv.org/abs/2406.05031} {arXiv:2406.05031 [hep-ph]} \BibitemShut {NoStop}%
\bibitem [{\citenamefont {Alcubierre}\ \emph {et~al.}(2022)\citenamefont {Alcubierre}, \citenamefont {Barranco}, \citenamefont {Bernal}, \citenamefont {Degollado}, \citenamefont {Diez-Tejedor}, \citenamefont {Jaramillo}, \citenamefont {Megevand}, \citenamefont {N\'u\~nez},\ and\ \citenamefont {Sarbach}}]{Alcubierre:2021psa}%
  \BibitemOpen
  \bibfield  {author} {\bibinfo {author} {\bibfnamefont {Miguel}\ \bibnamefont {Alcubierre}}, \bibinfo {author} {\bibfnamefont {Juan}\ \bibnamefont {Barranco}}, \bibinfo {author} {\bibfnamefont {Argelia}\ \bibnamefont {Bernal}}, \bibinfo {author} {\bibfnamefont {Juan~Carlos}\ \bibnamefont {Degollado}}, \bibinfo {author} {\bibfnamefont {Alberto}\ \bibnamefont {Diez-Tejedor}}, \bibinfo {author} {\bibfnamefont {V\'\i{}ctor}\ \bibnamefont {Jaramillo}}, \bibinfo {author} {\bibfnamefont {Miguel}\ \bibnamefont {Megevand}}, \bibinfo {author} {\bibfnamefont {Dar\'\i{}o}\ \bibnamefont {N\'u\~nez}}, \ and\ \bibinfo {author} {\bibfnamefont {Olivier}\ \bibnamefont {Sarbach}},\ }\bibfield  {title} {\enquote {\bibinfo {title} {{Extreme \ensuremath{\ell}-boson stars}},}\ }\href {\doibase 10.1088/1361-6382/ac5fc2} {\bibfield  {journal} {\bibinfo  {journal} {Class. Quant. Grav.}\ }\textbf {\bibinfo {volume} {39}},\ \bibinfo {pages} {094001} (\bibinfo {year} {2022})},\ \Eprint {http://arxiv.org/abs/2112.04529} {arXiv:2112.04529
  [gr-qc]} \BibitemShut {NoStop}%
\bibitem [{\citenamefont {Jaramillo}\ \emph {et~al.}(2023)\citenamefont {Jaramillo}, \citenamefont {Mart\'\i{}nez-Carbajal}, \citenamefont {Degollado},\ and\ \citenamefont {N\'u\~nez}}]{Jaramillo:2023lgk}%
  \BibitemOpen
  \bibfield  {author} {\bibinfo {author} {\bibfnamefont {V\'\i{}ctor}\ \bibnamefont {Jaramillo}}, \bibinfo {author} {\bibfnamefont {Daniel}\ \bibnamefont {Mart\'\i{}nez-Carbajal}}, \bibinfo {author} {\bibfnamefont {Juan~Carlos}\ \bibnamefont {Degollado}}, \ and\ \bibinfo {author} {\bibfnamefont {Dar\'\i{}o}\ \bibnamefont {N\'u\~nez}},\ }\bibfield  {title} {\enquote {\bibinfo {title} {{Born-Infeld boson stars}},}\ }\href {\doibase 10.1088/1475-7516/2023/07/017} {\bibfield  {journal} {\bibinfo  {journal} {JCAP}\ }\textbf {\bibinfo {volume} {07}},\ \bibinfo {pages} {017} (\bibinfo {year} {2023})},\ \Eprint {http://arxiv.org/abs/2303.13666} {arXiv:2303.13666 [gr-qc]} \BibitemShut {NoStop}%
\bibitem [{\citenamefont {Kleihaus}\ \emph {et~al.}(2009)\citenamefont {Kleihaus}, \citenamefont {Kunz}, \citenamefont {Lammerzahl},\ and\ \citenamefont {List}}]{Kleihaus:2009kr}%
  \BibitemOpen
  \bibfield  {author} {\bibinfo {author} {\bibfnamefont {Burkhard}\ \bibnamefont {Kleihaus}}, \bibinfo {author} {\bibfnamefont {Jutta}\ \bibnamefont {Kunz}}, \bibinfo {author} {\bibfnamefont {Claus}\ \bibnamefont {Lammerzahl}}, \ and\ \bibinfo {author} {\bibfnamefont {Meike}\ \bibnamefont {List}},\ }\bibfield  {title} {\enquote {\bibinfo {title} {{Charged Boson Stars and Black Holes}},}\ }\href {\doibase 10.1016/j.physletb.2009.03.066} {\bibfield  {journal} {\bibinfo  {journal} {Phys. Lett. B}\ }\textbf {\bibinfo {volume} {675}},\ \bibinfo {pages} {102--115} (\bibinfo {year} {2009})},\ \Eprint {http://arxiv.org/abs/0902.4799} {arXiv:0902.4799 [gr-qc]} \BibitemShut {NoStop}%
\bibitem [{\citenamefont {Kumar}\ \emph {et~al.}(2014)\citenamefont {Kumar}, \citenamefont {Kulshreshtha},\ and\ \citenamefont {Shankar~Kulshreshtha}}]{Kumar:2014kna}%
  \BibitemOpen
  \bibfield  {author} {\bibinfo {author} {\bibfnamefont {Sanjeev}\ \bibnamefont {Kumar}}, \bibinfo {author} {\bibfnamefont {Usha}\ \bibnamefont {Kulshreshtha}}, \ and\ \bibinfo {author} {\bibfnamefont {Daya}\ \bibnamefont {Shankar~Kulshreshtha}},\ }\bibfield  {title} {\enquote {\bibinfo {title} {{Boson stars in a theory of complex scalar fields coupled to the U(1) gauge field and gravity}},}\ }\href {\doibase 10.1088/0264-9381/31/16/167001} {\bibfield  {journal} {\bibinfo  {journal} {Class. Quant. Grav.}\ }\textbf {\bibinfo {volume} {31}},\ \bibinfo {pages} {167001} (\bibinfo {year} {2014})},\ \Eprint {http://arxiv.org/abs/1605.07210} {arXiv:1605.07210 [hep-th]} \BibitemShut {NoStop}%
\bibitem [{\citenamefont {{Ashtekar}}\ and\ \citenamefont {{Magnon-Ashtekar}}(1979)}]{1979JMP....20..793A}%
  \BibitemOpen
  \bibfield  {author} {\bibinfo {author} {\bibfnamefont {Abhay}\ \bibnamefont {{Ashtekar}}}\ and\ \bibinfo {author} {\bibfnamefont {Anne}\ \bibnamefont {{Magnon-Ashtekar}}},\ }\bibfield  {title} {\enquote {\bibinfo {title} {{On conserved quantities in general relativity}},}\ }\href {\doibase 10.1063/1.524151} {\bibfield  {journal} {\bibinfo  {journal} {Journal of Mathematical Physics}\ }\textbf {\bibinfo {volume} {20}},\ \bibinfo {pages} {793--800} (\bibinfo {year} {1979})}\BibitemShut {NoStop}%
\bibitem [{\citenamefont {Gourgoulhon}(2010)}]{Gourgoulhon:2010ju}%
  \BibitemOpen
  \bibfield  {author} {\bibinfo {author} {\bibfnamefont {Eric}\ \bibnamefont {Gourgoulhon}},\ }\bibfield  {title} {\enquote {\bibinfo {title} {{An Introduction to the theory of rotating relativistic stars}},}\ }in\ \href@noop {} {\emph {\bibinfo {booktitle} {{CompStar 2010: School and Workshop on Computational Tools for Compact Star Astrophysics}}}}\ (\bibinfo {year} {2010})\ \Eprint {http://arxiv.org/abs/1003.5015} {arXiv:1003.5015 [gr-qc]} \BibitemShut {NoStop}%
\bibitem [{\citenamefont {Niikura}\ \emph {et~al.}(2019)\citenamefont {Niikura} \emph {et~al.}}]{Niikura:2017zjd}%
  \BibitemOpen
  \bibfield  {author} {\bibinfo {author} {\bibfnamefont {Hiroko}\ \bibnamefont {Niikura}} \emph {et~al.},\ }\bibfield  {title} {\enquote {\bibinfo {title} {{Microlensing constraints on primordial black holes with Subaru/HSC Andromeda observations}},}\ }\href {\doibase 10.1038/s41550-019-0723-1} {\bibfield  {journal} {\bibinfo  {journal} {Nature Astron.}\ }\textbf {\bibinfo {volume} {3}},\ \bibinfo {pages} {524--534} (\bibinfo {year} {2019})},\ \Eprint {http://arxiv.org/abs/1701.02151} {arXiv:1701.02151 [astro-ph.CO]} \BibitemShut {NoStop}%
\bibitem [{\citenamefont {Smyth}\ \emph {et~al.}(2020)\citenamefont {Smyth}, \citenamefont {Profumo}, \citenamefont {English}, \citenamefont {Jeltema}, \citenamefont {McKinnon},\ and\ \citenamefont {Guhathakurta}}]{Smyth:2019whb}%
  \BibitemOpen
  \bibfield  {author} {\bibinfo {author} {\bibfnamefont {Nolan}\ \bibnamefont {Smyth}}, \bibinfo {author} {\bibfnamefont {Stefano}\ \bibnamefont {Profumo}}, \bibinfo {author} {\bibfnamefont {Samuel}\ \bibnamefont {English}}, \bibinfo {author} {\bibfnamefont {Tesla}\ \bibnamefont {Jeltema}}, \bibinfo {author} {\bibfnamefont {Kevin}\ \bibnamefont {McKinnon}}, \ and\ \bibinfo {author} {\bibfnamefont {Puragra}\ \bibnamefont {Guhathakurta}},\ }\bibfield  {title} {\enquote {\bibinfo {title} {{Updated Constraints on Asteroid-Mass Primordial Black Holes as Dark Matter}},}\ }\href {\doibase 10.1103/PhysRevD.101.063005} {\bibfield  {journal} {\bibinfo  {journal} {Phys. Rev. D}\ }\textbf {\bibinfo {volume} {101}},\ \bibinfo {pages} {063005} (\bibinfo {year} {2020})},\ \Eprint {http://arxiv.org/abs/1910.01285} {arXiv:1910.01285 [astro-ph.CO]} \BibitemShut {NoStop}%
\bibitem [{\citenamefont {Wyrzykowski}\ \emph {et~al.}(2011)\citenamefont {Wyrzykowski} \emph {et~al.}}]{Wyrzykowski:2011tr}%
  \BibitemOpen
  \bibfield  {author} {\bibinfo {author} {\bibfnamefont {L.}~\bibnamefont {Wyrzykowski}} \emph {et~al.},\ }\bibfield  {title} {\enquote {\bibinfo {title} {{The OGLE View of Microlensing towards the Magellanic Clouds. IV. OGLE-III SMC Data and Final Conclusions on MACHOs}},}\ }\href {\doibase 10.1111/j.1365-2966.2011.19243.x} {\bibfield  {journal} {\bibinfo  {journal} {Mon. Not. Roy. Astron. Soc.}\ }\textbf {\bibinfo {volume} {416}},\ \bibinfo {pages} {2949} (\bibinfo {year} {2011})},\ \Eprint {http://arxiv.org/abs/1106.2925} {arXiv:1106.2925 [astro-ph.GA]} \BibitemShut {NoStop}%
\bibitem [{\citenamefont {Blaineau}\ \emph {et~al.}(2022)\citenamefont {Blaineau} \emph {et~al.}}]{Blaineau:2022nhy}%
  \BibitemOpen
  \bibfield  {author} {\bibinfo {author} {\bibfnamefont {T.}~\bibnamefont {Blaineau}} \emph {et~al.},\ }\bibfield  {title} {\enquote {\bibinfo {title} {{New limits from microlensing on Galactic black holes in the mass range 10 M\ensuremath{\odot} \ensuremath{<} M \ensuremath{<} 1000 M\ensuremath{\odot}}},}\ }\href {\doibase 10.1051/0004-6361/202243430} {\bibfield  {journal} {\bibinfo  {journal} {Astron. Astrophys.}\ }\textbf {\bibinfo {volume} {664}},\ \bibinfo {pages} {A106} (\bibinfo {year} {2022})},\ \Eprint {http://arxiv.org/abs/2202.13819} {arXiv:2202.13819 [astro-ph.GA]} \BibitemShut {NoStop}%
\bibitem [{\citenamefont {Mr{\'o}z}\ \emph {et~al.}(2024{\natexlab{a}})\citenamefont {Mr{\'o}z} \emph {et~al.}}]{Mroz:2024mse}%
  \BibitemOpen
  \bibfield  {author} {\bibinfo {author} {\bibfnamefont {Przemek}\ \bibnamefont {Mr{\'o}z}} \emph {et~al.},\ }\bibfield  {title} {\enquote {\bibinfo {title} {{No massive black holes in the Milky Way halo}},}\ }\href {\doibase 10.1038/s41586-024-07704-6} {\bibfield  {journal} {\bibinfo  {journal} {Nature}\ }\textbf {\bibinfo {volume} {632}},\ \bibinfo {pages} {749--751} (\bibinfo {year} {2024}{\natexlab{a}})},\ \Eprint {http://arxiv.org/abs/2403.02386} {arXiv:2403.02386 [astro-ph.GA]} \BibitemShut {NoStop}%
\bibitem [{\citenamefont {Mr{\'o}z}\ \emph {et~al.}(2024{\natexlab{b}})\citenamefont {Mr{\'o}z} \emph {et~al.}}]{Mroz:2024wia}%
  \BibitemOpen
  \bibfield  {author} {\bibinfo {author} {\bibfnamefont {Przemek}\ \bibnamefont {Mr{\'o}z}} \emph {et~al.},\ }\bibfield  {title} {\enquote {\bibinfo {title} {{Limits on Planetary-mass Primordial Black Holes from the OGLE High-cadence Survey of the Magellanic Clouds}},}\ }\href {\doibase 10.3847/2041-8213/ad8e68} {\bibfield  {journal} {\bibinfo  {journal} {Astrophys. J. Lett.}\ }\textbf {\bibinfo {volume} {976}},\ \bibinfo {pages} {L19} (\bibinfo {year} {2024}{\natexlab{b}})},\ \Eprint {http://arxiv.org/abs/2410.06251} {arXiv:2410.06251 [astro-ph.CO]} \BibitemShut {NoStop}%
\bibitem [{\citenamefont {Kaup}(1968)}]{Kaup:1968zz}%
  \BibitemOpen
  \bibfield  {author} {\bibinfo {author} {\bibfnamefont {David~J.}\ \bibnamefont {Kaup}},\ }\bibfield  {title} {\enquote {\bibinfo {title} {{Klein-Gordon Geon}},}\ }\href {\doibase 10.1103/PhysRev.172.1331} {\bibfield  {journal} {\bibinfo  {journal} {Phys. Rev.}\ }\textbf {\bibinfo {volume} {172}},\ \bibinfo {pages} {1331--1342} (\bibinfo {year} {1968})}\BibitemShut {NoStop}%
\bibitem [{\citenamefont {Clesse}\ and\ \citenamefont {Garc{\'\i}a-Bellido}(2017)}]{Clesse:2016vqa}%
  \BibitemOpen
  \bibfield  {author} {\bibinfo {author} {\bibfnamefont {Sebastien}\ \bibnamefont {Clesse}}\ and\ \bibinfo {author} {\bibfnamefont {Juan}\ \bibnamefont {Garc{\'\i}a-Bellido}},\ }\bibfield  {title} {\enquote {\bibinfo {title} {{The clustering of massive Primordial Black Holes as Dark Matter: measuring their mass distribution with Advanced LIGO}},}\ }\href {\doibase 10.1016/j.dark.2016.10.002} {\bibfield  {journal} {\bibinfo  {journal} {Phys. Dark Univ.}\ }\textbf {\bibinfo {volume} {15}},\ \bibinfo {pages} {142--147} (\bibinfo {year} {2017})},\ \Eprint {http://arxiv.org/abs/1603.05234} {arXiv:1603.05234 [astro-ph.CO]} \BibitemShut {NoStop}%
\bibitem [{\citenamefont {Carr}\ \emph {et~al.}(2016)\citenamefont {Carr}, \citenamefont {Kuhnel},\ and\ \citenamefont {Sandstad}}]{Carr:2016drx}%
  \BibitemOpen
  \bibfield  {author} {\bibinfo {author} {\bibfnamefont {Bernard}\ \bibnamefont {Carr}}, \bibinfo {author} {\bibfnamefont {Florian}\ \bibnamefont {Kuhnel}}, \ and\ \bibinfo {author} {\bibfnamefont {Marit}\ \bibnamefont {Sandstad}},\ }\bibfield  {title} {\enquote {\bibinfo {title} {{Primordial Black Holes as Dark Matter}},}\ }\href {\doibase 10.1103/PhysRevD.94.083504} {\bibfield  {journal} {\bibinfo  {journal} {Phys. Rev. D}\ }\textbf {\bibinfo {volume} {94}},\ \bibinfo {pages} {083504} (\bibinfo {year} {2016})},\ \Eprint {http://arxiv.org/abs/1607.06077} {arXiv:1607.06077 [astro-ph.CO]} \BibitemShut {NoStop}%
\bibitem [{\citenamefont {Giudice}\ \emph {et~al.}(2016)\citenamefont {Giudice}, \citenamefont {McCullough},\ and\ \citenamefont {Urbano}}]{Giudice:2016zpa}%
  \BibitemOpen
  \bibfield  {author} {\bibinfo {author} {\bibfnamefont {Gian~F.}\ \bibnamefont {Giudice}}, \bibinfo {author} {\bibfnamefont {Matthew}\ \bibnamefont {McCullough}}, \ and\ \bibinfo {author} {\bibfnamefont {Alfredo}\ \bibnamefont {Urbano}},\ }\bibfield  {title} {\enquote {\bibinfo {title} {{Hunting for Dark Particles with Gravitational Waves}},}\ }\href {\doibase 10.1088/1475-7516/2016/10/001} {\bibfield  {journal} {\bibinfo  {journal} {JCAP}\ }\textbf {\bibinfo {volume} {10}},\ \bibinfo {pages} {001} (\bibinfo {year} {2016})},\ \Eprint {http://arxiv.org/abs/1605.01209} {arXiv:1605.01209 [hep-ph]} \BibitemShut {NoStop}%
\bibitem [{\citenamefont {Jaramillo}\ \emph {et~al.}(2022)\citenamefont {Jaramillo}, \citenamefont {Sanchis-Gual}, \citenamefont {Barranco}, \citenamefont {Bernal}, \citenamefont {Degollado}, \citenamefont {Herdeiro}, \citenamefont {Megevand},\ and\ \citenamefont {N\'u\~nez}}]{Jaramillo:2022zwg}%
  \BibitemOpen
  \bibfield  {author} {\bibinfo {author} {\bibfnamefont {V\'\i{}ctor}\ \bibnamefont {Jaramillo}}, \bibinfo {author} {\bibfnamefont {Nicolas}\ \bibnamefont {Sanchis-Gual}}, \bibinfo {author} {\bibfnamefont {Juan}\ \bibnamefont {Barranco}}, \bibinfo {author} {\bibfnamefont {Argelia}\ \bibnamefont {Bernal}}, \bibinfo {author} {\bibfnamefont {Juan~Carlos}\ \bibnamefont {Degollado}}, \bibinfo {author} {\bibfnamefont {Carlos}\ \bibnamefont {Herdeiro}}, \bibinfo {author} {\bibfnamefont {Miguel}\ \bibnamefont {Megevand}}, \ and\ \bibinfo {author} {\bibfnamefont {Dar\'\i{}o}\ \bibnamefont {N\'u\~nez}},\ }\bibfield  {title} {\enquote {\bibinfo {title} {{Head-on collisions of \ensuremath{\ell}-boson stars}},}\ }\href {\doibase 10.1103/PhysRevD.105.104057} {\bibfield  {journal} {\bibinfo  {journal} {Phys. Rev. D}\ }\textbf {\bibinfo {volume} {105}},\ \bibinfo {pages} {104057} (\bibinfo {year} {2022})},\ \Eprint {http://arxiv.org/abs/2202.00696} {arXiv:2202.00696 [gr-qc]} \BibitemShut {NoStop}%
\bibitem [{\citenamefont {Lira}\ \emph {et~al.}(2025)\citenamefont {Lira}, \citenamefont {Villegas}, \citenamefont {Antelis}, \citenamefont {Jaramillo}, \citenamefont {Moreno},\ and\ \citenamefont {N\'u\~nez}}]{Lira:2024cma}%
  \BibitemOpen
  \bibfield  {author} {\bibinfo {author} {\bibfnamefont {Mariana}\ \bibnamefont {Lira}}, \bibinfo {author} {\bibfnamefont {Laura~O.}\ \bibnamefont {Villegas}}, \bibinfo {author} {\bibfnamefont {Javier~M.}\ \bibnamefont {Antelis}}, \bibinfo {author} {\bibfnamefont {V\'\i{}ctor}\ \bibnamefont {Jaramillo}}, \bibinfo {author} {\bibfnamefont {Claudia}\ \bibnamefont {Moreno}}, \ and\ \bibinfo {author} {\bibfnamefont {Dar\'\i{}o}\ \bibnamefont {N\'u\~nez}},\ }\bibfield  {title} {\enquote {\bibinfo {title} {{On the detectability of gravitational waves emitted from head-on collisions of $\ell $-boson stars}},}\ }\href {\doibase 10.1007/s10714-025-03397-y} {\bibfield  {journal} {\bibinfo  {journal} {Gen. Rel. Grav.}\ }\textbf {\bibinfo {volume} {57}},\ \bibinfo {pages} {68} (\bibinfo {year} {2025})},\ \Eprint {http://arxiv.org/abs/2411.19401} {arXiv:2411.19401 [gr-qc]} \BibitemShut {NoStop}%
\bibitem [{\citenamefont {Herrera-Aguilar}\ and\ \citenamefont {Nucamendi}(2015)}]{Herrera-Aguilar:2015kea}%
  \BibitemOpen
  \bibfield  {author} {\bibinfo {author} {\bibfnamefont {Alfredo}\ \bibnamefont {Herrera-Aguilar}}\ and\ \bibinfo {author} {\bibfnamefont {Ulises}\ \bibnamefont {Nucamendi}},\ }\bibfield  {title} {\enquote {\bibinfo {title} {{Kerr black hole parameters in terms of the redshift/blueshift of photons emitted by geodesic particles}},}\ }\href {\doibase 10.1103/PhysRevD.92.045024} {\bibfield  {journal} {\bibinfo  {journal} {Phys. Rev. D}\ }\textbf {\bibinfo {volume} {92}},\ \bibinfo {pages} {045024} (\bibinfo {year} {2015})},\ \Eprint {http://arxiv.org/abs/1506.05182} {arXiv:1506.05182 [gr-qc]} \BibitemShut {NoStop}%
\bibitem [{\citenamefont {Wang}\ \emph {et~al.}(2022)\citenamefont {Wang}, \citenamefont {Lee},\ and\ \citenamefont {Lin}}]{Wang:2022ouq}%
  \BibitemOpen
  \bibfield  {author} {\bibinfo {author} {\bibfnamefont {Chen-Yu}\ \bibnamefont {Wang}}, \bibinfo {author} {\bibfnamefont {Da-Shin}\ \bibnamefont {Lee}}, \ and\ \bibinfo {author} {\bibfnamefont {Chi-Yong}\ \bibnamefont {Lin}},\ }\bibfield  {title} {\enquote {\bibinfo {title} {{Null and timelike geodesics in the Kerr-Newman black hole exterior}},}\ }\href {\doibase 10.1103/PhysRevD.106.084048} {\bibfield  {journal} {\bibinfo  {journal} {Phys. Rev. D}\ }\textbf {\bibinfo {volume} {106}},\ \bibinfo {pages} {084048} (\bibinfo {year} {2022})},\ \Eprint {http://arxiv.org/abs/2208.11906} {arXiv:2208.11906 [gr-qc]} \BibitemShut {NoStop}%
\bibitem [{\citenamefont {Villegas}\ \emph {et~al.}(2023)\citenamefont {Villegas}, \citenamefont {Ramirez-Codiz}, \citenamefont {Jaramillo}, \citenamefont {Degollado}, \citenamefont {Moreno}, \citenamefont {N\'u\~nez},\ and\ \citenamefont {Romero-Cruz}}]{Villegas:2022jzt}%
  \BibitemOpen
  \bibfield  {author} {\bibinfo {author} {\bibfnamefont {Laura~O.}\ \bibnamefont {Villegas}}, \bibinfo {author} {\bibfnamefont {Eduardo}\ \bibnamefont {Ramirez-Codiz}}, \bibinfo {author} {\bibfnamefont {V\'\i{}ctor}\ \bibnamefont {Jaramillo}}, \bibinfo {author} {\bibfnamefont {Juan~Carlos}\ \bibnamefont {Degollado}}, \bibinfo {author} {\bibfnamefont {Claudia}\ \bibnamefont {Moreno}}, \bibinfo {author} {\bibfnamefont {Dar\'\i{}o}\ \bibnamefont {N\'u\~nez}}, \ and\ \bibinfo {author} {\bibfnamefont {Fernando~J.}\ \bibnamefont {Romero-Cruz}},\ }\bibfield  {title} {\enquote {\bibinfo {title} {{Determination of the angular momentum of the Kerr black hole from equatorial geodesic motion}},}\ }\href {\doibase 10.1088/1475-7516/2023/08/007} {\bibfield  {journal} {\bibinfo  {journal} {JCAP}\ }\textbf {\bibinfo {volume} {08}},\ \bibinfo {pages} {007} (\bibinfo {year} {2023})},\ \Eprint {http://arxiv.org/abs/2211.10464} {arXiv:2211.10464 [gr-qc]} \BibitemShut {NoStop}%
\end{thebibliography}%
\end{document}